\documentclass[12pt,a4paper]{article}
\usepackage{amssymb}
\usepackage{epsfig}
\usepackage{color}

\def\s2w{\sin^2 \theta_W}

\def\be{\begin{equation}}
\def\ee{\end{equation}}
\def\ba{\begin{eqnarray}}
\def\ea{\end{eqnarray}}

\def\w{\wedge}
\def\d{{\rm d}}

\def\r{\rho}
\def\a{\alpha}

\def\b{\beta}

\def\g{\gamma}
\def\G{\Gamma}

\def\D{\Delta}
\def\e{\epsilon}

\def\m{\mu}
\def\n{\nu}

\def\l{\lambda}
\def\L{\Lambda}
\def\s{\sigma}
\def\S{\Sigma}

\def\t{\theta}

\def\IR{\relax{\rm I\kern-.18em R}}

\def\tr{\,{\rm tr}\,}

\def\a{\alpha}
\def\b{\beta}
\def\g{\gamma}
\def\G{\Gamma}
\def\d{{\rm d}}

\def\e{\epsilon}

\def\m{\mu}
\def\n{\nu}
\def\r{\rho}
\def\l{\lambda}

\def\s{\sigma}
\def\o{\omega}

\def\O{\Omega}

\def\ks{{k \kern-.5em /}}
\def\es{{\e \kern-.4em /}}
\def\ds{{\partial \kern-.5em /}}
\def\Ds{{D \kern-.6em /}}

\def\gh{{\hat g}}

\def\hN{\hat{N}}

\begin{document}
%
%
%
%
\begin{titlepage}
\begin{flushright}
LPTENS-05/11\\
LMU-ASC 23/05\\
hep-th/0503173
\end{flushright}
\vskip 2.5cm

\begin{center}{\Large\bf Special geometry of local Calabi-Yau manifolds} \\
\vspace{2mm} {\Large\bf and superpotentials}
\vspace{2mm} {\Large \bf from holomorphic matrix models}
\vskip 1.5cm {\bf Adel Bilal$^{1}$ and Steffen
Metzger$^{1,2}$}

\vskip.3cm $^1$ Laboratoire de Physique Th\'eorique,
\'Ecole Normale Sup\'erieure - CNRS\\
24 rue Lhomond, 75231 Paris Cedex 05, France

\vskip.3cm $^2$ Arnold-Sommerfeld-Center for Theoretical
Physics,\\
Department f\"ur Physik, Ludwig-Maximilians-Universit\"at\\
Theresienstr. 37, 80333 Munich, Germany\\

\vskip.3cm {\small e-mail: {\tt adel.bilal@lpt.ens.fr,
metzger@physique.ens.fr}}
\end{center}
\vskip .5cm

\begin{center}
{\bf Abstract}
\end{center}
\begin{quote}
We analyse the (rigid) special geometry of a class of local
Calabi-Yau manifolds given by hypersurfaces in $\mathbb{C}^4$ as
$W'(x)^2+f_0(x)+v^2+w^2+z^2=0$, that arise in the study of the
large $N$ duals of four-dimensional $\mathcal{N}=1$ supersymmetric
$SU(N)$ Yang-Mills theories with adjoint field $\Phi$ and
superpotential $W(\Phi)$. The special geometry relations are
deduced from the planar limit of the corresponding holomorphic
matrix model. The set of cycles is split into a bulk sector, for
which we obtain the standard rigid special geometry relations, and
a set of relative cycles, that come from the non-compactness of
the manifold, for which we find cut-off dependent corrections to
the usual special geometry relations. The (cut-off independent)
prepotential is identified with the (analytically continued) free
energy of the holomorphic matrix model in the planar limit. On the
way, we clarify various subtleties pertaining to the saddle point
approximation of the holomorphic matrix model. A formula for the
superpotential of IIB string theory with background fluxes on
these local Calabi-Yau manifolds is proposed that is based on
pairings similar to the ones of relative cohomology.
\end{quote}

\end{titlepage}
\setcounter{footnote}{0}
\setlength{\baselineskip}{.7cm}
\newpage
%
%
\newtheorem{proposition}{Proposition}[section]
\newtheorem{theorem}[proposition]{Theorem}
\newtheorem{definition}[proposition]{Definition}
\newtheorem{conjecture}[proposition]{Conjecture}

\section{Introduction\label{Intro}}
Compactifications of string theory on Calabi-Yau manifolds have
been studied for almost two decades \cite{CHSW85}. One
particularly appealing property of Calabi-Yau compactifications is
that the special geometry structure of the effective supergravity
theory \cite{dWvP} can be understood from the fact that the
(K\"ahler and complex structure) moduli spaces of a Calabi-Yau
manifold are special K\"ahler manifolds. What is more, the
prepotential that corresponds to the complex structure
deformations can be expressed in terms of period integrals on the
Calabi-Yau space \cite{CO90}. For these periods one can deduce the
Picard-Fuchs differential equations and so get interesting
physical quantities from the solutions of these equations. In
general the computation of the K\"ahler part is more complicated
as in this case the period integrals are corrected by world-sheet
instantons. However, mirror symmetry comes to the rescue since the
K\"ahler moduli space of a Calabi-Yau $X$ can be mapped to the
complex structure moduli space of its mirror $\tilde X$. The
K\"ahler prepotential can then be computed using the mirror map.

To be more precise, if $\mathcal{F}$ is the prepotential on the
moduli space of complex structures of a Calabi-Yau
manifold $X$, one has the special geometry relations \cite{CO90}\\
\parbox{14cm}{
\begin{eqnarray}
X^I&=&\int_{\G_{A^I}}\O\ ,\nonumber\\
\mathcal{F}_I\equiv{\partial\mathcal{F}\over\partial
X^I}&=&\int_{\G_{B_I}}\O\ .\nonumber
\end{eqnarray}}\hfill\parbox{8mm}{\begin{eqnarray}\label{SG}\end{eqnarray}}\\
Here $\O$ is the unique holomorphic $(3,0)$-form on the
Calabi-Yau, $I\in\{0,\ldots ,h^{2,1}\}$ and
$\{\G_{A^I},\G_{B_J}\}$ is a symplectic basis of $H_3(X)$. The
homogeneous function $\mathcal{F}$ can be obtained from
$2\mathcal{F}=X^I\mathcal{F}_I$.

Triggered by the success of Seiberg and Witten in solving
four-dimensional gauge theories \cite{SW94} it became apparent
that {\it local} Calabi-Yau manifolds are also quite useful to
extract important information about four-dimensional physics
\cite{KLMVW96}. These are {\it non-compact} K\"ahler manifolds
with vanishing first Chern class and the idea is that they are
local models of proper Calabi-Yau manifolds. They also appeared in
the context of geometrical engineering and local mirror symmetry,
see e.g. \cite{KKV96}, where they are constructed from toric
geometry, and especially the analysis of the topological string
with these target manifolds has led to a variety of interesting
results \cite{AKMV03, ADKMV03}, see \cite{M04} for a review.

More recently a different class of local Calabi-Yau manifolds
appeared \cite{KKLM99}, \cite{CIV01}. These are given as
deformations of the space
$\mathcal{O}(-2)\oplus\mathcal{O}(0)\rightarrow \mathbb{CP}^1$ by
a polynomial $W$. Instead of a single $\mathbb{CP}^1$, the
deformed manifold, let us call it $Y$, contains $n$ two-spheres if
the degree of $W$ is $n+1$. These spaces can be taken through a
geometric transition \cite{CIV01}, similar to the conifold
transition of Gopakumar and Vafa \cite{GV98}. The resulting space,
we call it $X$, is a hypersurface in $\mathbb{C}^4$ described by
\begin{equation}
W'(x)^2+f_0(x)+v^2+w^2+z^2=0\ ,\label{deformed}
\end{equation}
where $(v,w,x,z)\in\mathbb{C}^4$ and $f_0(x)$ is a polynomial of degree $n-1$.\\
An obvious and important question to ask is whether we can find
special geometry for these manifolds as well. In fact, a local
Calabi-Yau manifold also comes with a holomorphic $(3,0)$-form
$\O$ and we want to check whether its integrals over an
appropriate basis of three-cycles satisfy (\ref{SG}). Clearly, the
naive special geometry relations need to be modified since our
local Calabi-Yau manifold $X$ now contains a non-compact
three-cycle $\G_{\hat B}$ and the integral of $\O$ over this cycle
is divergent. This can be remedied by introducing a cut-off
$\L_0$, but then the integral over the regulated cycle is cut-off
dependent whereas the prepotential is expected to be independent
of any cut-off. The question therefore is how the relation
\begin{equation}
\int_{\G_{B_I}}\O\stackrel{?}{=}{\partial\mathcal{F}\over\partial
X^I}
\end{equation}
should be modified to make sense on local Calabi-Yau manifolds.
Related issues have been addressed recently in \cite{DOV04}.

Mathematically the relation between the spaces $Y$ and $X$ is
obvious from the fact that both are related to the singular space
\begin{equation}
W'(x)^2+v^2+w^2+z^2=0\ .\label{singular}
\end{equation}
$X$ is simply the deformation\footnote{Usually we use the word
``deformation'' in an intuitive sense, but what is meant here is
the precise mathematical term.} and $Y$ is nothing but the small
resolution of all the singularities in (\ref{singular}). Following
previous work in \cite{GV98,V01} it was shown in \cite{CIV01} that
this geometric transition has a beautiful physical interpretation
in type IIB string theory. Starting from the manifold $Y$ one can
wrap $N_i$ D5-branes around the $i$-th $\mathbb{CP}^1$ to get an
effective $\mathcal{N}=1$ $U(N)$ theory with an adjoint field
$\Phi$ in a vacuum that breaks the gauge group as $U(N)\rightarrow
U(N_1)\times\ldots\times U(N_{n})$, where $N:=\sum_iN_i$. After
the transition the branes disappear and we are left with a dual
$\mathcal{N}=1$ $U(1)^{n}$ theory with background flux three-form
$H$ with $\int_{\G_{A^i}}H=N_i$. The effective superpotential of
the dual gauge theory can be calculated from the formula
\cite{CIV01}
\begin{equation}
W_{eff}=\sum_{i=1}^{n}\left(\int_{\G_{A^i}}H\int_{\G_{B_i}}\O-\int_{\G_{B_i}}H\int_{\G_{A^i}}\O\right)\
.\label{Wintro}
\end{equation}
The $\G_{A^i},\G_{B_i}$ form a symplectic basis of three-cycles
and Eq. (\ref{Wintro}) obviously is invariant under symplectic
changes of basis which include the ``electric magnetic" duality
transformations $\G_{A^i}\rightarrow\G_{B_i},\
\G_{B_i}\rightarrow-\G_{A^i}$. Note that we do not write the
right-hand side as $\int_X H\w\O$ as it is not clear whether the
Riemann bilinear relation can be extended to non-compact
Calabi-Yau manifolds without modification. In \cite{CIV01} the
$\G_A$-cycles where taken to be compact and the $\G_B$-cycles all
non-compact. But $\int_{\G_{B_i}}\O$ contains a term that diverges
polynomially together with a term with a logarithmic divergence.
The latter has a nice interpretation in terms of the $\b$-function
of the gauge theory but the polynomial divergence has not been
understood. One of the goals of this note is to shed some light on
this aspect.

In a series of influential papers \cite{DV02}, Dijkgraaf and Vafa
reviewed these local Calabi-Yau manifolds and showed that the
field theory corresponding to branes wrapped on $\mathbb{CP}^1$s
in $Y$, which is holomorphic Chern-Simons theory \cite{Wi95},
reduces to a {\it holomorphic} matrix model. In fact, as will be
discussed below, the structure of the space (\ref{deformed}) is
essentially captured by a Riemann surface and a very similar
Riemann surface appears in the planar limit of the matrix model.
This is why one can learn something about the local Calabi-Yau
manifold from an analysis of the well-understood matrix model.
Specifically we are interested in understanding the detailed form
of the special geometry relations on local Calabi-Yau manifolds
from the analysis in the holomorphic matrix model.

The holomorphic matrix model is similar to the hermitian one, but
its potential $W(x)$ is defined on the complex plane, has complex
coefficients and the integration is performed over complex
$\hN\times \hN$ matrices with eigenvalues that are constrained to
lie on some path $\g$ in the complex plane. The precise definition
and solution involve various subtleties, many of which have been
addressed in \cite{La03}, and others will be clarified in this
note. The planar limit of the free energy of the matrix model is
given, as usual, by a saddle point approximation. We show that
saddle point solutions exist only for an appropriate choice of the
path $\g$, which is determined self-consistently in such a way
that {\it all} critical points of $W(x)$ appear as {\it stable}
critical points along the path! For the case of finite $\hN$ this
can be seen from an approximate solution of the saddle point
equations. In the planar limit one usually constructs the
eigenvalue density $\r_0(s)$ ($s$ is a real parameter along $\g$)
from the Riemann surface that appears in this limit. As a matter
of fact, every Riemann surface that arises in the planar limit of
a matrix model leads to a {\it real} density $\r_0(s)$. A way to
see this is to note that the filling fractions, i.e. the numbers
of eigenvalues in certain domains of $\mathbb{C}$, can be
calculated as real integrals over $\r_0(s)$. One can also turn the
argument around and construct a $\r_0(s)$ from an arbitrary
hyperelliptic Riemann surface. In general this $\r_0(s)$ will be
complex and one obtains constraints on the moduli of the surface
from the condition that $\r_0(s)$ should be real. Once we fix the
filling fractions, the moduli of the Riemann surface are in fact
uniquely determined. This, in turn, gives the positions of the
cuts $\mathcal{C}_i$ which support the eigenvalues and as the cuts
have to lie on the path $\g$ we get conditions
for the path.\\
Coupling the filling fractions to sources then gives a planar free
energy $F_0(J_i)$, and its Legendre transform
$\mathcal{F}_0(\tilde S_i)$ is the candidate prepotential. In
fact, the $\tilde S_i$ are related to the filling fractions in
such a way that they are given by the period integrals over the
(compact) $\a_i$-cycles on the Riemann surface and the
${\partial\mathcal{F}_0\over\partial \tilde S_i}(\tilde S_j)$ can
be shown to be integrals over the corresponding (compact)
$\b_i$-cycles. These properties can in fact be generalised to
arbitrary hyperelliptic Riemann surfaces by analytically
continuing the $\tilde S_i$ to complex values. The prepotential
then still has the same form, but now it depends on complex
variables (and then it can no longer be interpreted as the planar
limit of the free energy of a matrix model). This proves the
standard special geometry relations for the standard cycles. The
same methods allow us to derive the modifications of the special
geometry relations for the relative cycles appearing in the setup.
Indeed, the non-compact period integrals contain, in addition to
the derivatives of the prepotential, a polynomial and a
logarithmical cut-off dependence and can therefore {\it not} be
written as a derivative of the prepotential. While the logarithmic
divergence is interpreted as related to the $\b$-function of the
dual gauge theory, the polynomial divergence has no counterpart
and should not appear in the effective superpotential. This will
be achieved by defining appropriate pairings similar to the ones
appearing in relative cohomology.

The analysis in the matrix model and the derivation of the special
geometry relations show that it is useful to work with a
symplectic basis of (relative) one-cycles on the Riemann surface
which consists of $n-1$ compact cycles $\a^i$ and the $n-1$
corresponding compact cycles $\b_i$, together with two (relative)
cycles $\hat\a$ and $\hat\b$, where only $\hat\b$ is non-compact.
Indeed, then one can perform symplectic transformations in the set
$\{\a^i,\b_j\}$ maintaining the usual special geometry relations.
However, once the relative cycle $\hat\b$ is combined with other
cycles the special geometry relations are modified. Quite
importantly the transformed prepotential always stays finite and
cut-off independent.

This paper is organised as follows. In the next section we explain
the structure of the local Calabi-Yau spaces we are considering.
In particular we review how the set of three-cycles in $X$ maps to
the set of relative one-cycles on a Riemann surface with marked
points. Section three deals with holomorphic matrix models, where
the potential of the model is chosen to correspond to the $W(x)$
of the Calabi-Yau manifold. We start with a short exposition of
general facts from holomorphic matrix models and then discuss how
to deal with the above-mentioned subtleties. We explain how
special geometry arises from the matrix model and how the
modifications for the non-compact cycles can be derived.
Furthermore, we discuss the properties of the prepotential and how
electric-magnetic duality is implemented. In section four we
propose a formula for the effective superpotential of IIB string
theory on these local Calabi-Yau manifolds. It contains the
above-mentioned pairings that are similar to the ones appearing in
relative cohomology and provides a precise reformulation of the
formulae found in \cite{CIV01} and \cite{DV02}. Section five
contains our conclusions.

\section{Local Calabi-Yau Manifolds and hyperelliptic Riemann surfaces}
\setcounter{equation}{0} Let then $(v,w,x,z)\in\mathbb{C}^4$,
$W(x)$ a polynomial of degree $n+1$ with $W(0)=g_0$, leading
coefficient one, and non-degenerate critical points, i.e. if
$W'(p)=0$ then $W''(p)\neq0$. Furthermore let $f_0(x)$ a
polynomial of degree $n-1$. In this note we are only interested in
local Calabi-Yau manifolds $X$ described by the equation
\begin{equation}
F(v,w,x,z)\equiv W'(x)^2+f_0(x)+v^2+w^2+z^2=0\ .\label{locCY}
\end{equation}
In particular, we want to see how the special geometry relations
(\ref{SG}) have to be modified in this case.

The holomorphic three-form on $X$ is given as\footnote{If
$F(v,w,x,z)$ is a holomorphic function on $\mathbb{C}^4$ then $\d
F$ is perpendicular to the hypersurface $F=0$. From the
holomorphic four-form $\d v\w\d w\w\d x\w\d z$ on $\mathbb{C}^4$
one defines the holomorphic three-form $\O$ on $F=0$ as the form
that satisfies $\d v\w\d w\w\d x\w\d z=\O\w\d F$.} \cite{CIV01},
\cite{AGLV}
\begin{equation}
\O={\d v\w\d w\w\d x\over{2z} }\ .
\end{equation}
Because of the simple dependence of the surface (\ref{locCY}) on
$v$ and $w$, every three-cycle of the space (\ref{locCY}) can be
understood \cite{KLMVW96} as a fibration of a two-sphere over a
line segment in the hyperelliptic Riemann surface $\S$,
\begin{equation}
y^2=W'(x)^2+f_0(x)=\prod_{i=1}^{n}(x-a_i^+)(x-a_i^-)\
,\label{Riemannsurface}
\end{equation}
of genus $\gh=n-1$, see \cite{L96} for a review. $\S$ is a
two-sheeted covering of the complex plane where the two sheets are
connected by $n$ cuts between the points $a_i^-$ and $a_i^+$. Our
conventions are such that if $y_0$ is the branch of the Riemann
surface with $y_0(x)\sim W'(x)$ for $|x|\rightarrow \infty$, then
$y_0$ is defined on the upper sheet and $y_1=-y_0$ on the lower
one. For compact three-cycles the line segment connects two of the
branch points of the curve and the volume of the $S^2$-fibre
depends on the position on the base line segment. At the end
points of the segment one has $y^2=0$ and the volume of the sphere
shrinks to zero size. Non-compact three-cycles on the other hand
are fibrations of $S^2$ over a half-line that runs from one of the
branch points to infinity on the Riemann surface. Integration over
the fibre is elementary and gives
\begin{equation}\label{fibration}
\int_{S^2}\O=\pm 2\pi i\ y(x)\d x\ ,
\end{equation}
(the sign ambiguity will be fixed momentarily) and thus the
integral of the holomorphic $\O$ over a three-cycle is reduced to
an integral of $\pm 2\pi i y\d x$ over a line segment in $\S$.
Clearly, the integrals over line segments that connect two branch
points can be rewritten in terms of integrals over compact cycles
on the Riemann surface, whereas the integrals over non-compact
three-cycles can be expressed as integrals over a line that links
the two infinities on the two complex sheets. In fact, the
one-form
\begin{equation}
\zeta:=y\d x
\end{equation}
is meromorphic and diverges at infinity (poles of order $n+2$) on
the two sheets and therefore it is well-defined only on the
Riemann surface with the two infinities, we denote them by $Q$ and
$Q'$, removed. This surface with two points cut out is called
$\hat\S$. We are naturally led to consider the relative
homology\footnote{Let $M$ be a manifold and $N$ a submanifold of
$M$ and $C_j(M),C_j(N)$ the set of $j$-chains in $M$ and $N$,
respectively. One defines the group $C_j(M,N)$ of equivalence
classes of $j$-chains $c_j\in C_j(M)$, where $[c_j]:=c_j+C_j(N)$.
Then two chains are equivalent if they differ only by a chain in
$N$. As usual $H_j(M,N)=Z_j(M,N)/B_j(M,N)$, where
$Z_j(M,N):=\{[c_j]\in C_j(M,N):\partial[c_j]=[0]\}$ and
$B_j(M,N):=\partial C_{j+1}(M,N)$. Note that a representative
$c_j$ of an element in $H_j(M,N)$ may have a boundary as long as
the boundary lies in $N$.} $H_1(\S,\{Q,Q'\})$. This group contains
both, all the compact cycles, as well as cycles connecting $Q$ and
$Q'$ on the Riemann surface. If we take, for example,
$W(x)={x^2\over2}$ and $f_0(x)=-\m$ the surface (\ref{locCY}) is
nothing but the deformed conifold, which is $T^*S^3$. This space
contains two three-cycles, the compact base $\G_{\hat\a}\cong
S^3$, which maps to the compact one-cycle $\hat\a$ surrounding the
cut of the surface $y^2=x^2-\m$, and the non-compact fiber
$\G_{\hat\b}:=T_p^*S^3$, which maps to the non-compact one-cycle
$\hat\b$ which runs from $Q'$, i.e. infinity on the lower sheet
through the cut to $Q$, i.e. infinity on the upper sheet. This can
be generalised readily for arbitrary polynomials $W,f_0$ and one
finds a one-to-one correspondence between the (compact and
non-compact) three-cycles in
(\ref{locCY}) and $H_1(\S,\{Q,Q'\})$.\\
\begin{figure}[h]
\centering
\includegraphics[width=1.0\textwidth]{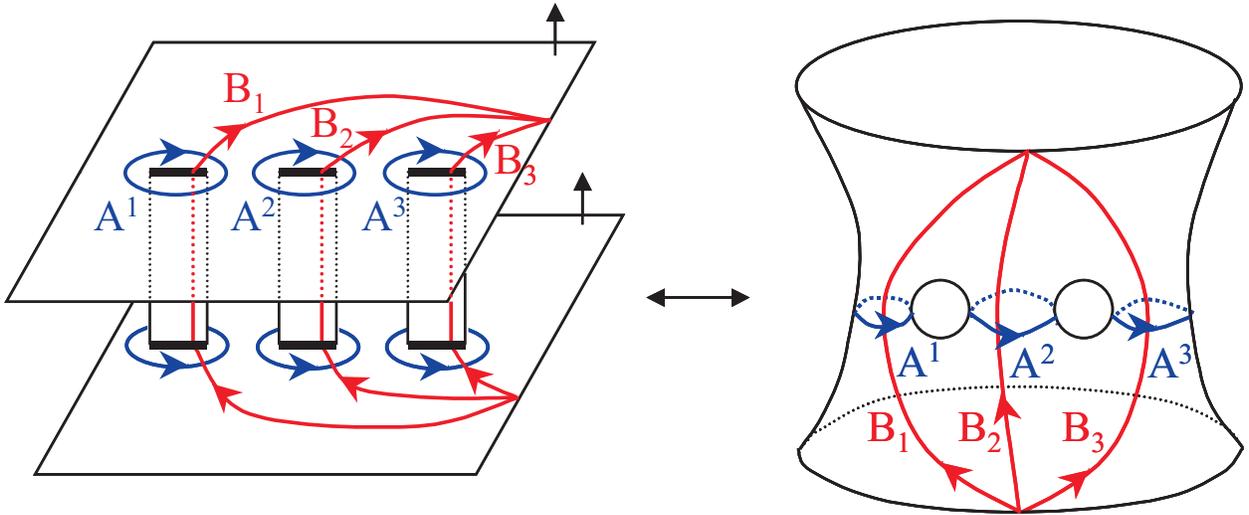}\\
\caption[]{A symplectic choice of compact $A$- and non-compact
$B$-cycles for $n=3$. Note that the orientation of the two planes
on the left-hand side is chosen such that both normal vectors
point to the top. This is why the orientation of the $A$-cycles is
different on the two planes. To go from the representation of the
Riemann surface on the left to the one on the right one has to
flip the upper plane.} \label{ABcycles}
\end{figure}
There are various symplectic bases of this relative homology
group. One such basis is $\{A^i, B_j\}$, with $i,j=1,\ldots n$,
where the one-cycle $A^i$ runs around the $i$-th cut and the
relative one-cycles $B_j$ are all non-compact and run from $Q'$
through the $j$-th cut to $Q$. This is the choice of cycles used
in $\cite{CIV01}$ and it is shown in Fig.\ref{ABcycles}.

\begin{figure}[h]
\centering
\includegraphics[width=1.0\textwidth]{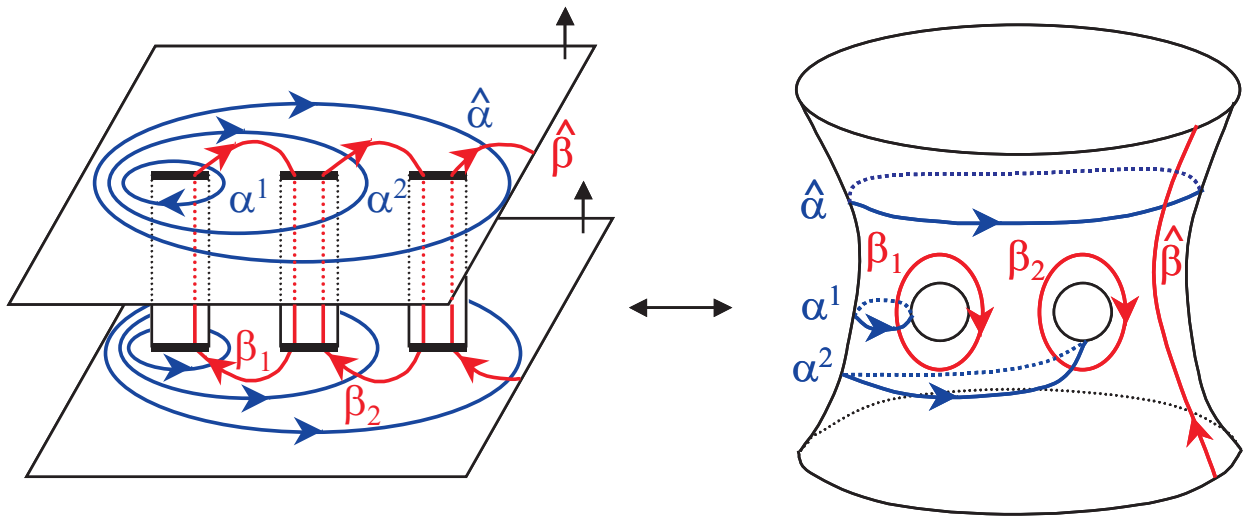}\\
\caption[]{A symplectic set of cycles containing only one
non-compact cycle $\hat\b$. The cycle $\a^i$ surrounds $i$ of the
cuts, whereas the cycle $\b_i$ runs from cut $i$ to cut $i+1$ on
the upper sheet and from cut $i+1$ to cut $i$ on the lower one. As
before one has to flip the upper plane to find the representation
of the surface on the right.} \label{albecycles}
\end{figure}

Another useful symplectic basis is the set
$\{\a^i,\b_j,\hat\a,\hat\b\}$, with $\gh=n-1$ compact cycles
$\a^i$ and $\gh=n-1$ compact cycles $\b_i$, with intersection
numbers $\a^i\cap\b_j=\delta^i_j$, together with one compact cycle
$\hat\a$ and one non-compact cycle $\hat\b$, with
$\hat\a\cap\hat\b=1$, see Fig.\ref{albecycles}. Note that although
these bases are equivalent, since one can be obtained from the
other by a symplectic transformation, the second basis is much
more useful for our purpose. This is because it contains only one
non-compact cycle and the new features coming from the
non-compactness of the space should be contained entirely in the
corresponding integral. Finally, we take $\G_{A^i},\G_{B_j}$ to be
the $S^2$-fibrations over $A^i,B_j$ and $\G_{\a^i},\G_{\b_j}$
$S^2$-fibrations over $\a^i,\b_j$.

So the problem effectively reduces to calculating the
integrals\footnote{The sign ambiguity of (\ref{fibration}) has now
been fixed, since we have made specific choices for the
orientation of the cycles. Furthermore, we use the (standard)
convention that the cut of $\sqrt{x}$ is along the negative real
axis of the complex $x$-plane. Also, on the right-hand side we
used that the integral of $\zeta$ over the line segment is $1\over
2$ times the integral over a closed cycle $\g$.}
\begin{equation}
\int_{\G_{\g}}\O=-i\pi\int_{\g}\zeta\ \ \ \mbox{for}\ \ \
\g\in\{\a^i,\b_j,\hat\a,\hat\b\}\ .\label{integrals}
\end{equation}
For $\gh=1$ they can be reduced to various combinations of
elliptic integrals of the three kinds.

As we mentioned already, we expect new features to be contained in
the integral $\int_{\hat\b}\zeta$, where $\hat\b$ runs from $Q'$
on the lower sheet to $Q$ on the upper one. Indeed, it is easy to
see that this integral is divergent. It will be part of our task
to understand and properly treat this divergence. As usual, we
will regulate the integral and we have to make sure that physical
quantities do not depend on the regulator and remain finite once
the regulator is removed. Usually this is achieved by simply
discarding the divergent part. Instead, we want to give a more
intrinsic geometric prescription that will be similar to standard
procedures in relative cohomology. To render the integral finite
we simply cut out two ``small'' discs around the points $Q,Q'$. If
$x,x'$ are coordinates on the upper and lower sheet respectively,
we restrict ourselves to $|x|\leq \L_0$, $|x'|\leq\L_0$,
$\L_0\in\mathbb{R}$. Furthermore we take the cycle $\hat\b$ to run
from the point $\L_0'$ on the real axis of the lower sheet to
$\L_0$ on the real axis of the upper sheet. (Actually we could
take $\L_0$ and $\L_0'$ to be complex. We will come back to this
point later on.)

\section{Holomorphic Matrix Models and Special Geometry}
\setcounter{equation}{0}

Our goal is to relate the integrals (\ref{integrals}) to the
prepotential $\mathcal{F}_0$. It turns out that in order to
address this problem it is useful to perform calculations in the
matrix model that corresponds to our local Calabi-Yau manifold.
Indeed, the analysis of Dijkgraaf and Vafa tells us \cite{DV02}
that one should identify the prepotential and the planar limit of
the free energy of the holomorphic matrix model with potential
$W(x)$. Therefore, our goal will be to find the special geometry
relation in the holomorphic matrix model and to see how the
integrals (\ref{integrals}) over the cycles
$\a^i,\b_j,\hat\a,\hat\b$ are related to the planar limit of the
free energy.\\
One should note, however, that in the matrix model the filling fractions $S_i$, related to the
integrals over the $A$-cycles, are manifestly real, even though
$W(x)$ has arbitrary complex coefficients. So, strictly speaking,
the matrix model does {\it not} explore the full moduli space of
the Calabi-Yau manifold. Nevertheless, we will see that all
relevant formulae can be immediately continued to complex values
of the $S_i$ and, in particular, the
special geometry relations continue to be true.

\subsection{The Holomorphic Matrix Model}
The proper definition of the holomorphic matrix model is somewhat
more subtle than the one of the hermitian matrix model. Many of
these subtleties were nicely addressed in \cite{La03} and we will
briefly review them here. The usual identification of the planar
limit with the saddle point approximation involves even more
subtleties which we will have to clarify in this subsection.
Particular attention is paid to the dependence of the free energy
on the various parameters.

\subsubsection{The partition function and convergence properties}
We begin by defining the partition function of the {\it
holomorphic} one-matrix model following \cite{La03}. In order to
do so, one chooses a smooth path
$\g:\mathbb{R}\rightarrow\mathbb{C}$ without self-intersection,
such that $\dot{\g}(u)\neq 0\ \forall u\in\mathbb{R}$ and
$|\g(u)|\rightarrow \infty$ for $u\rightarrow\pm\infty$. Consider
the ensemble $\G(\g)$ of\footnote{We reserve the letter $N$ for
the number of colours in a $U(N)$ gauge theory. It is important to
distinguish between $N$ in the gauge theory and $\hat{N}$ in the
matrix model.} $\hN\times\hN$ complex matrices $M$ with spectrum
spec$(M)=\{\l_1,\ldots\l_{\hN}\}$ in\footnote{Here and in the
following we will write $\g$ for both the function and its image.}
$\g$ and distinct eigenvalues,
\begin{equation}
\Gamma(\gamma):=\{M\in\mathbb{C}^{\hN\times\hN}:\mbox{spec}(M)\subset\gamma,\
\mbox{all}\ \l_m \ \mbox{distinct}\}\ .
\end{equation}
The holomorphic measure on $\mathbb{C}^{\hat{N}\times \hat{N}}$ is
just $\d M\equiv\wedge_{p,q}\d M_{pq}$ with some appropriate sign
convention. The (super-)potential is
\begin{equation}
W(x):=g_0+\sum_{k=1}^{n+1} {g_k\over k} x^k,\ \ \ g_{n+1}=1\ .
\end{equation}
Without loss of generality we have chosen $g_{n+1}=1$. The only
restriction for the other complex parameters $\{g_k\}_{k=0,\ldots
n}$, collectively denoted by $g$, comes from the fact that the $n$
critical points $\m_i$ of $W$ should not be degenerate, i.e.
$W''(\m_i)\neq0$ if $W'(x)=\prod_{i=1}^n(x-\m_i)$. Then the
partition function of the holomorphic one-matrix model is
\begin{equation}
Z(\Gamma(\g),g,g_s,\hN):=C_{\hN}\int_{\Gamma(\g)}\d M \
\exp\left(-{1\over g_s} \tr W(M)\right),
\end{equation}
where $g_s$ is a positive coupling constant and $C_{\hN}$ is some
normalisation factor. To avoid cluttering the notation we will
omit the dependence on $\g$ and $g$ and write
$Z(g_s,\hN):=Z(\G(\g),g,g_s,\hN)$. As usual one diagonalises $M$
and performs the integral over the diagonalising matrices. The
constant $C_{\hN}$ is chosen in such a way that one arrives at
\begin{equation}
Z(g_s,\hN)={1\over{\hat{N}!}}\int_\gamma\d\lambda_1\ldots\int_\gamma\d\lambda_{\hat{N}}
\exp\left({-\hN^2S(g_s,\hN;\l_m)}\right)=:e^{-F(g_s,\hN)}\ ,
\end{equation}
where
\begin{equation}
S(g_s,\hN;\l_m)={1\over\hN^2g_s}\sum_{m=1}^{\hN}W(\l_m)-{1\over\hN^2}\sum_{p\neq
q}\ln(\l_p-\l_q)\ .
\end{equation}
See \cite{La03} for more details.

\begin{figure}[h]
\centering
\includegraphics{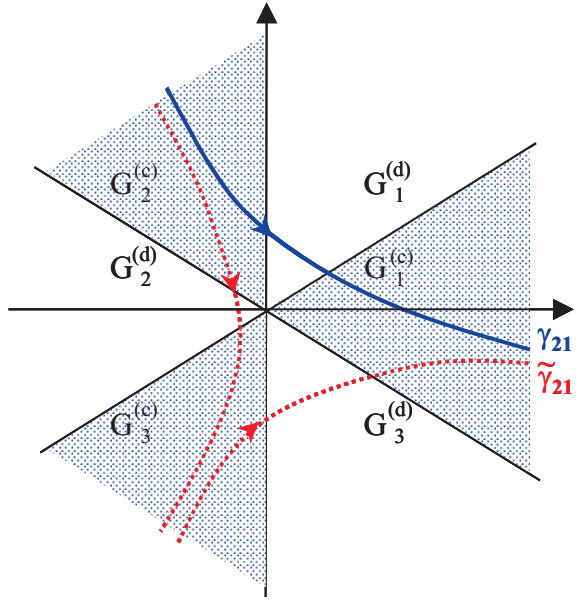}\\
\caption[]{Example of convergence and divergence domains for $n=2$
and a possible choice of $\g_{21}$. Because of holomorphicity the
path can be deformed without changing the partition function, for
instance one could use the path $\tilde\g_{21}$ instead.}
\label{convergence1}
\end{figure}

The convergence of the $\l_m$ integrals depends on the polynomial
$W$ and the choice of the path $\g$. For given $W$ the asymptotic
part of the complex plane ($|x|$ large) can be divided into
convergence domains $G_k^{(c)}$ and divergence domains
$G_k^{(d)}$, $k=1,\ldots n+1$, where $e^{-{1\over g_s}W(x)}$
converges, respectively diverges as $|x|\rightarrow\infty$. The
path $\g$ has to be chosen \cite{La03} to go from some convergence
domain $G_k^{(c)}$ to some other $G_l^{(c)}$, with $k\neq l$; call
such a path $\g_{kl}$, see Fig.\ref{convergence1}.
Then the value of the partition function depends only on the pair
$(k,l)$ and, because of holomorphicity, is not sensitive to
deformations of $\g_{kl}$. In particular, instead of $\g_{kl}$ we
can make the equivalent choice
\begin{equation}
\tilde\g_{kl}=\g_{p_1p_2}\cup\g_{p_2p_3}\cup\ldots\cup\g_{p_{n-1}p_{n}}\cup\g_{p_{n}p_{n+1}}\
\ \ \mbox{with}\ p_1=k,\ p_{n+1}=l ,\label{tildegamma}
\end{equation}
as shown in Fig.\ref{convergence1}. Here we split the path into
$n$ components, each component running from one convergence
sector to another. Again, due to holomorphicity we can choose the
decomposition in such a way that every component $\g_{p_ip_{i+1}}$
runs through one of the $n$ critical points of $W$ in
$\mathbb{C}$, or at least comes close to it. This choice of
$\tilde\g_{kl}$ will turn out to be very useful to understand the
saddle point approximation discussed
below.
Hence, the partition function and the free energy depend on the
pair $(k,l),g,g_s$ and $\hN$. Of course, one can always relate the
partition function for arbitrary $(k,l)$ to one with $(k',1)$, $k'=k-l+1$ mod
$n$, and
redefined coupling constants $g$.

\subsubsection{Matrix model technology}
Next, we need to recall some standard technology adapted to the
holomorphic matrix model. We first assume that the path $\g$
consists of a single connected piece. The case (\ref{tildegamma})
will be discussed later on. Let $s$ be the length coordinate of
the path $\g$, centered at some point on $\g$, and let $\l(s)$
denote the parameterisation of $\g$ with respect to this
coordinate. Then, for an eigenvalue $\l_m$ on $\g$, one has
$\l_m=\l(s_m)$ and the partition function can be rewritten as
\begin{equation}\label{part(s)}
Z(g_s,\hN)={1\over{\hN!}}\int_{\mathbb{R}}\d
s_1\ldots\int_{\mathbb{R}}\d
s_{\hat{N}}\prod_{l=1}^{\hat{N}}\dot{\l}(s_l)\exp\left(-\hN^2S(g_s,\hN;\l(s_m))\right)\
.
\end{equation}
The spectral density is defined as
\begin{equation}
\r(s,s_m):={1\over \hat{N}}\sum_{m=1}^{\hat{N}}\delta(s-s_m)\ ,
\end{equation}
so that $\r$ is normalised to one,
$\int_{-\infty}^\infty\r(s,s_m)\d s=1$. The normalised trace of
the resolvent of the matrix $M$ is given by
\begin{equation}\label{omrho}
\o (x,s_m):={1\over \hat{N}}\tr {1\over {x-M}}={1\over
\hat{N}}\sum_{m=1}^{\hat{N}}{1\over {x-\l(s_m)}}=\int\d
s{\r(s,s_m)\over {x-\l(s)}}\ ,
\end{equation}
for $x\in\mathbb{C}$. Following \cite{La03} we decompose the
complex plane into domains $D_i$, $i\in\{1,\ldots, n\}$, with
mutually disjoint interior, $( \cup_i\overline{D}_i=\mathbb{C},\ \
D_i\cap D_j=\emptyset\ \ \mbox{for}\ i\neq j)$. These domains are
chosen in such a way that $\g$ intersects each $\overline{D}_i$
along a single line segment $\D_i$, and $\cup_i\D_i=\g$.
Furthermore, $\m_i$, the $i$-th critical point of $W$, should lie
in the interior of $D_i$. One defines
\begin{equation}
\chi_i(M):=\int_{\partial D_i} {\d x\over 2\pi i}{1\over{x-M}}\ ,
\end{equation}
(which projects on the space spanned by the eigenvectors of $M$
whose eigenvalues lie in $D_i$), and the filling fractions
$\tilde\s_i(\l_m):={1\over\hN}\tr\chi_i(M)$ and
\begin{equation}
\s_i(s_m):=\tilde\s_i(\l(s_m))=\int\d s\
\r(s,s_m)\chi_i(\l(s))=\int_{\partial D_i}{\d x\over 2\pi i}\
\o(x,s_m)\ ,\label{fillingfraction}
\end{equation}
(which count the eigenvalues in the domain $D_i$, times $1/\hN$). Obviously
\begin{equation}\label{sumnu}
\sum_{i=1}^{n}\s_i(s_m)=1\ .
\end{equation}

One can apply standard methods (e.g. the ones of \cite{K99}) to
derive the loop equations of the holomorphic matrix model,
\begin{equation}\label{loop}
\langle \o(x,s_m)^2\rangle-{1\over
t}W'(x)\langle\o(x,s_m)\rangle-{1\over 4t^2}\langle
f(x,s_m)\rangle=0\ .
\end{equation}
Here
\begin{equation}
t=g_s\hN\label{t}
\end{equation}
is the quantity that will be held fixed in the planar limit below,
\begin{equation}
f(x,s_m):=-{4t\over{\hN}}\sum_{m=1}^{\hat{N}}{{W'(x)-W'(\l(s_m))}\over{x-\l(s_m)}}=-4t\int
\d s\ \r(s,s_m){{W'(x)-W'(\l(s))}\over{x-\l(s)}}\ ,
\end{equation}
and the expectation value is defined for a $h(\l_m)=h(\l(s_m))$ as
usual:
\begin{equation}
\langle h(\l_m)\rangle:={1\over Z(g_s,\hN)}\cdot{1\over
\hN!}\int_{\g}\d\l_1\ldots\int_{\g}\d\l_{\hat{N}}\ h(\l_m)
\exp\left(-\hN^2S(g_s,\hN;\l_m)\right)\ .
\end{equation}

\begin{figure}[h]
\centering
\includegraphics{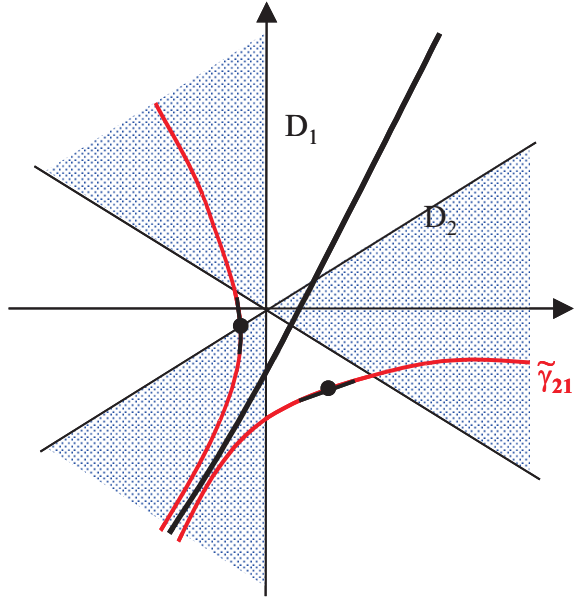}\\
\caption[]{For the cubic potential of Fig.\ref{convergence1} we
show the choice of the domains $D_1$ and $D_2$ and of the path
$\tilde\g_{21}$ with respect to the two critical points, as well
as the cuts that form around these points.} \label{curvesinD}
\end{figure}

It will be useful to define an effective action as
\begin{eqnarray}\label{effaction}
&&S_{eff}(g_s,\hN;s_m):=S(g_s,\hN;\l(s_m))-{1\over\hN^2}\sum_{m=1}^{\hN}\ln(\dot\l(s_m))\nonumber\\
&&=\int\d s\ \r(s;s_m)\left({1\over
t}W(\l(s))-{1\over{\hN}}\ln(\dot{\l}(s))-\mathcal{P}\int\d s' \
\r(s';s_p)\ln(\l(s)-\l(s'))\right)\ \ \ \ \ \ \ \ \ \
\end{eqnarray}
so that
\begin{equation}
Z(g_s,\hN)={1\over {\hN!}}\int\d s_1\ldots\int\d s_{\hat{N}}\
\exp\left(-\hN^2 S_{eff}(g_s,\hN;s_m)\right)\ .
\end{equation}
Note that the principal value is defined as
\begin{equation}\label{Pvalue}
\mathcal{P}\
\ln\left(\l(s)-\l(s')\right)={1\over2}\lim_{\e\rightarrow0}\left[\ln\left(\l(s)-\l(s')+
i\e\dot\l(s)\right)+\ln\left(\l(s)-\l(s')-i\e\dot\l(s)\right)\right]\
.
\end{equation}
The equations of motion corresponding to this effective action,
${\delta S_{eff}\over\delta s_m}=0$, read
\begin{equation}\label{eommatrix}
{1\over t}W'(\l(s_m))={2\over{\hat{N}}}\sum_{p=1,\ p\neq
m}^{\hN}{1\over{\l(s_m)-\l(s_p)}}+{1\over{\hat{N}}}{\ddot{\l}(s_m)\over\dot{\l}(s_m)^2}\
.
\end{equation}
Using these equations of motion one can show that
\begin{eqnarray}\label{eomloop}
&&\o(x,s_m)^2-{1\over t}W'(x)\o(x,s_m)-{1\over 4 t^2}f(x,s_m)+\nonumber\\
&&+{1\over\hN}{\d\over\d x}\o(x,s_m)+{1\over\hN^2}
\sum_{m=1}^{\hN}{\ddot\l(s_m)\over\dot\l(s_m)^2}{1\over{x-\l(s_m)}}=0\
.
\end{eqnarray}
\\
\underline{Solutions of the equations of motion}\\
Note that in general the effective action is a complex function of
the real $s_m$. Hence, in general, i.e. for a generic path
$\g_{kl}$ with parameterisation $\l(s)$, there will be no solution
to (\ref{eommatrix}). One clearly expects that the existence of
solutions must constrain the path $\l(s)$ appropriately. Let us
study this in more detail.\\
Recall that we defined the domains
$D_i$ in such a way that $\m_i\subset D_i$. Let $\hN_i$ be the
number of eigenvalues $\l(s_m)$ which lie in the domain $D_i$, so
that $\sum_{i=1}^n\hN_i=\hN$, and denote them by $\l(s_a^{(i)})$,
$a\in\{1,\ldots\hN_i\}$.\\
Solving the equations of motion in general is a formidable
problem. To get a good idea, however, recall the picture of
$\hN_i$ fermions filled into the $i$-th ``minimum" of ${1\over
t}W$ \cite{K91}. For small $t$ the potential is deep and the
fermions are located not too far from the minimum, in other words
all the eigenvalues are close to $\m_i$. To be more precise
consider (\ref{eommatrix}) and drop the last term, an
approximation that will be justified momentarily. Let us take $t$
to be small and look for solutions\footnote{One might try the
general ansatz $\l(s_a^{(i)})=\m_i+\e\delta\l_a^{(i)}$ but it
turns out that a solution can be found only if $\e\sim\sqrt{t}$.}
$\l(s_a^{(i)})=\m_i+\sqrt{t}\delta\l_a^{(i)}$, where
$\delta\l_a^{(i)}$ is of order one. So, we assume that the
eigenvalues $\l(s_a^{(i)})$ are not too far from the critical
point $\m_i$. Then the equation reads
\begin{equation}
W''(\m_i)\delta\l_a^{(i)}={2\over\hN}\sum_{b=1,\ b\neq
a}^{\hN_i}{1\over{\delta\l^{(i)}_a-\delta\l_b^{(i)}}}+o(\sqrt{t})\
,
\end{equation}
so we effectively reduced the problem to finding the solution for
$n$ distinct quadratic potentials. If we set $z_a:=\sqrt{\hN
W''(\m_i)\over2}\delta\l_a^{(i)}$ and neglect the
$o(\sqrt{t})$-terms this gives
\begin{equation}
z_a=\sum_{b=1,\ b\neq a}^{\hN_i}{1\over {z_a-z_b}}\ ,
\end{equation}
which can be solved explicitly for small $\hN_i$. It is obvious
that $\sum_{a=1}^{\hN_i}z_a=0$, and one finds that there is a
unique solution (up to permutations) with the $z_a$ symmetrically
distributed around 0 on the real axis. This justifies a posteriori
that we really can neglect the term proportional to the second
derivative of $\l(s)$, at least to leading order. Furthermore,
setting $W''(\m_i)=|W''(\m_i)|e^{i\phi_i}$ one finds that the
$\l(s_a^{i})$ sit on a tilted line segment around $\m_i$ where the
angle of the tilt is given by $-\phi_i/2$. This means for example
that for a potential with $W'(x)=x(x-1)(x+1)$ the eigenvalues are
distributed on the real axis around $\pm1$ and on the imaginary
axis around 0. Note further that, in general, the reality of $z_a$
implies that
${W''(\m_i)\over2}\left({\delta\l^{(i)}_a}\right)^2>0$ which tells
us that, close to $\m_i$, $W(\l(s))-W(\m_i)$ is real with a {\it
minimum} at $\l(s)=\m_i$.\\
So we have found that the path $\g_{kl}$ has to go through the
critical points $\m_i$ with a tangent direction fixed by the phase
of the second derivative of $W$. On the other hand, we know that
the partition function does not depend on the form of the path
$\g_{kl}$. Of course, there is no contradiction: if one wants to
compute the partition function from a saddle point expansion, as
we will do below, and as is implicit in the planar limit, one has
to make sure that one expands around solutions of
(\ref{eommatrix}) and the existence of these solutions imposes
conditions on how to choose the path $\g_{kl}$. From now on we
will assume that the path is chosen in such a way that it
satisfies all these constraints. Furthermore, for later purposes
it will be useful to use the path $\tilde\g_{kl}$ of
(\ref{tildegamma}) chosen such that its part $\g_{p_ip_{i+1}}$
goes through all $\hN_i$ solutions $\l_a^{(i)}$,
$a=1,\ldots\hN_i$, and lies entirely in $D_i$, see
Fig.\ref{curvesinD}.
\\
It is natural to assume that these properties together with the
uniqueness of the solution (up to permutations) extend to higher
numbers of $\hN_i$ as well. Of course once one goes beyond the
leading order in $\sqrt{t}$ the eigenvalues are no longer
distributed on a straight line, but on a line segment that is
bent in general and that might or might not pass through $\m_i$.\\
\\
\underline{The large $\hN$ limit}\\
We are interested in the large $\hN$ limit of the matrix model
free energy. It is well known that the expectation values of the
relevant quantities like $\r$ or $\o$ have expansions of the form
\begin{equation}
\langle\r(s,s_m)\rangle=\sum_{I=0}^{\infty}\r_I(s)\hN^{-I}\ \ \ ,\
\ \ \langle\o(x,s_m)\rangle=\sum_{I=0}^{\infty}\o_I(x)\hN^{-I}\
.\label{omexp}
\end{equation}
Clearly, $\o_0(x)$ is related to $\r_0(s)$ by the large $\hN$
limit of (\ref{omrho}), namely
\begin{equation}\label{omegaofrho}
\o_0(x)=\int\d s{\r_0(s)\over{x-\l(s)}}\ .
\end{equation}
We saw already that an eigenvalue ensemble that solves the
equations of motion is distributed along line segments around the
critical points $\m_i$. In the limit $\hN\rightarrow \infty$ this
will turn into a continuous distribution on the segments
$\mathcal{C}_i$, through or close to the critical points of $W$.
Then $\r_0(s)$ has support only on these $\mathcal{C}_i$ and
$\o_0(x)$ is analytic in $\mathbb{C}$ with cuts $\mathcal{C}_i$.
Conversely, $\r_0(s)$ is given by the discontinuity of $\o_0(x)$
across its cuts:
\begin{equation}
\r_0(s):=\dot{\l}(s)\lim_{\e\rightarrow0}{1\over2\pi i}
[\o_0(\l(s)-i\e\dot{\l}(s))-\o_0(\l(s)+i\e\dot{\l}(s))]\label{solution}\
.
\end{equation}

The planar limit we are interested in is $\hN\rightarrow\infty$,
$g_s\rightarrow0$ with $t=g_s\hN$ held fixed. Hence we rewrite all
$\hN$ dependence as a $g_s$ dependence and consider the limit
$g_s\rightarrow 0$. Then, the equation of motion (\ref{eommatrix})
reduces to
\begin{equation}\label{eomplanar}
{1\over t}W'(\l(s))=2\mathcal{P}\int \d s'\
{\r_0(s')\over{\l(s)-\l(s')}}\ .
\end{equation}
Note that this equation is only valid for those $s$ where
eigenvalues exist, i.e. where $\r_0(s)\neq 0$. In principle one
can use this equation to compute the planar eigenvalue
distribution $\r_0(s)$ for given $W'$.\\
\\
\underline{Riemann surfaces and planar solutions}\\
The leading term in the expansion (\ref{omexp}) for
$\langle\o(s,s_m)\rangle$ can be calculated from a saddle point
approximation, where the $\{s_m\}$ are given by a solution
$\{s_m^*\}$ of (\ref{eommatrix}): $\o_0(x)=\o(x;s_m^*)$. This is
true for all ``microscopic'' operators, i.e. operators that do not
modify the saddle point equations (\ref{eommatrix}). (Things would
be different for ``macroscopic''operators like
$e^{\hN\sum_{p=1}^{\hN}V(\l_p)}$.) In particular, this shows that
expectation values factorise in the large $\hN$ limit, and the
loop equation (\ref{loop}) reduces to the algebraic equation
\begin{equation}\label{algconstraint}
\o_0(x)^2-{1\over t}W'(x)\o_0(x)-{1\over 4t^2}f_0(x)=0\ ,
\end{equation}
where
\begin{equation}\label{f0matrix}
f_0(x)=-4t\int \d s \ \r_0(s){{W'(x)-W'(\l(s))}\over{x-\l(s)}}
\end{equation}
is a polynomial of degree $n-1$ with leading coefficient $-4t$.
Note that this coincides with the planar limit of equation
(\ref{eomloop}). If we define
\begin{equation}\label{y0omega0}
y_0(x):=W'(x)-{2t}\o_0(x)\ ,
\end{equation}
then $y_0$ is one of the branches of the algebraic curve
\begin{equation}
y^2=W'(x)^2+f_0(x)\ ,\label{algcurve}
\end{equation}
as can be seen from (\ref{algconstraint}). On this curve we use
the same conventions as in section 2, i.e. $y_0(x)$ is defined on
the upper sheet and cycles and orientations are chosen as in
Fig.\ref{ABcycles} and Fig.\ref{albecycles}.

Solving a matrix model in the planar limit means to find a
normalised, real, non-negative $\r_0(s)$ and a path $\tilde
\g_{kl}$ which satisfy (\ref{omegaofrho}), (\ref{eomplanar}) and
(\ref{algconstraint}/\ref{f0matrix}) for a given potential $W(z)$
and a given asymptotics $(k,l)$ of $\g$.

Interestingly, for any algebraic curve (\ref{algcurve}) there is a
contour $\tilde \g_{kl}$ supporting a {\it formal} solution of the
matrix model in the planar limit. To construct it start from an
arbitrary polynomial $f_0(x)$ or order $n-1$, with leading
coefficient $-4t$, which is given together with the potential
$W(x)$ of order $n+1$. The corresponding Riemann surface is given
by (\ref{algcurve}), and we denote its branch points by $a_i^\pm$
and choose branch-cuts $\mathcal{C}_i$ between them. We can read
off the two solutions $y_0$ and $y_1=-y_0$ from (\ref{algcurve}),
where we take $y_0$ to be the one with a behaviour
$y_0\stackrel{x\rightarrow\infty}{\rightarrow}+W'(x)$. $\o_0(x)$
is defined as in (\ref{y0omega0}) and we choose a path
$\tilde\g_{kl}$ such that $\mathcal{C}_i\subset \tilde\g_{kl}$ for
all $i$. Then the formal planar spectral density satisfying all
the requirements can be determined from (\ref{solution}) (see
\cite{La03}). However, in general, this will lead to a {\it
complex} distribution $\r_0(s)$. This can be understood from the
fact the we constructed $\rho_0(s)$ from a completely arbitrary
hyperelliptic Riemann surface. However, in the matrix model the
algebraic curve (\ref{algcurve}) is {\it not} general, but the
coefficients $\a_k$ of $f_0(x)$ are constraint. This can be seen
by computing the filling fractions
\begin{equation}\label{nustar}
\n_i^*:=\langle\s_i(s_m)\rangle={1\over2\pi i}\int_{\partial
D_i}\o_0(x)\d x={1\over4\pi it}\int_{A^i}y_0(x)\d
x=\int_{\g^{-1}(\mathcal{C}_i)}\r_0(s)\d s\ ,
\end{equation}
which must be real and non-negative. Here we used the fact that
the $D_i$ were chosen such that $\g_{p_ip_{i+1}}\subset D_i$ and
therefore $\mathcal{C}_i\subset D_i$, so for $D_i$ on the upper
plane, $\partial D_i$ is homotopic to $-A^i$. Hence, ${\rm
Im}\left(i\int_{A^i}y(x)\d x\right)=0$ which constrains the
$\a_k$. We conclude that to construct distributions $\r_0(s)$ that
are relevant for the matrix model one can proceed along the lines
described above, but one has to impose the additional constraint
that $\r_0(s)$ is {\it real}. As for finite $\hN$, this will
impose conditions on the possible paths $\tilde\g_{kl}$ supporting
the eigenvalue distributions.

To see this, we assume that the coefficients $\a_k$ in $f_0(x)$
are small, so that the lengths of the cuts are small compared to
the distances between the different critical points:
$|a_i^+-a_i^-|\ll|\m_i-\m_j|$. Then in first approximation the
cuts are straight line segments between $a_i^+$ and $a_i^-$. For
$x$ close to the cut $\mathcal{C}_i$ we have
$y^2\approx(x-a_i^+)(x-a_i^-)\prod_{j\neq
i}(\m_i-\m_j)^2=(x-a_i^+)(x-a_i^-)(W''(\m_i))^2$. If we set
$W''(\m_i)=|W''(\m_i)|e^{i\phi_i}$ and
$a_i^+-a_i^-=r_ie^{i\psi_i}$, then, on the cut $\mathcal{C}_i$,
the path $\g$ is parameterised by
$\l(s)={a_i^+-a_i^-\over|a_i^+-a_i^-|}s=se^{i\psi_i}$, and we find
from (\ref{solution})
\begin{eqnarray}
\r_0(s)&=&{1\over2\pi t }\sqrt{|\l(s)-a_1^+|}\sqrt{|\l(s)-a_1^-|}\ |W''(\m_i)|e^{i(\phi_i+2\psi_i)}\nonumber\\
&=& {1\over2 \pi t }\sqrt{|\l(s)-a_1^+|}\sqrt{|\l(s)-a_1^-|}\
W''(\m_i)(\dot\l(s))^2\ .
\end{eqnarray}
So reality and positivity of $\r_0(s)$ lead to conditions on the
orientation of the cuts in the complex plane, i.e. on the path
$\g$:
\begin{equation}\label{condition}
\psi_i=-{\phi_i/2}\ \ \ \ \ \ ,\ \ \ \ \
W''(\m_i)(\dot\l(s))^2>0\ .
\end{equation}
These are precisely the conditions we already derived for the case
of finite $\hN$. We see that the two approaches are consistent
and, for given $W$ and fixed $\hN_i$ respectively $\n_i^*$, lead
to a unique\footnote{To be more precise the path $\tilde\g_{kl}$
is not entirely fixed. Rather, for every piece
$\tilde\g_{p_ip_{i+1}}$ we have the requirement that
$\mathcal{C}_i\subset\tilde\g_{p_ip_{i+1}}$.} solution
$\{\l(s),\r_0(s)\}$ with real and positive eigenvalue
distribution. Note again that the requirement of reality and
positivity of $\r_0(s)$ constrains the phases of $a_i^+-a_i^-$ and
hence the coefficients $\a_k$ of $f_0(x)$.

\subsubsection{The saddle point approximation for the partition function}
Recall that our goal is to find a relation between the planar
limit ($t=g_s\hN$ fixed, $g_s\rightarrow0$) of the free energy of
the matrix model and the period integrals of $y(x)\d x$ on the
corresponding Riemann curve. Since the standard planar free energy
$F_0(t)$ depends on $t$ only it cannot appear in relations like
(\ref{SG}), and one has to introduce a set of sources $J_i$ to
have a free energy that depends on more variables. In this
subsection we evaluate this source dependent free energy and its
Legendre transform $\mathcal{F}_0(t,S)$ in the planar limit using
a saddle point expansion.

We start by coupling sources to the filling
fractions,\footnote{Note that $\exp\left(-{\hN^2\over
t}\sum_{i=1}^{n-1}J_i\s_i(s_m)\right)$ looks like a macroscopic
operator that changes the equations of motion. However, because of
the special properties of $\s_i(s_m)$ we have
${\partial\over\partial s_n}\s_i(s_m)={1\over\hN}\int_{\partial
D_i}{\d x\over2\pi i}{\dot\l(s_n)\over(x-\l(s_n))^2}$. In
particular, for the path $\tilde \g_{kl}$ that will be chosen
momentarily and the corresponding domains $D_i$ the eigenvalues
$\l_m$ cannot lie on $\partial D_i$. Hence,
${\partial\over\partial s_n}\s_i(s_m)=0$ and the equations of
motion remain unchanged.}
\begin{eqnarray}\label{partsources}
Z(g_s,\hN,J)&:=&{1\over{\hN!}}\int_{\g}\d \l_1\ldots\int_{\g}\d
\l_{\hat{N}}\exp\left(-\hN^2S(g_s,\hN;\l_m)-{\hN\over
g_s}\sum_{i=1}^{n-1}J_i\tilde\s_i(\l_m)\right)\nonumber\\
&=&\exp\left(-F(g_s,\hN,J)\right)\ .
\end{eqnarray}
where $J:=\{J_1,\ldots,J_{n-1}\}$. Note that because of the
constraint $\sum_{i=1}^{n}\tilde\s_i(\l_m)=1$,
$\tilde\s_{n}(\l_m)$ is not an independent quantity and we can
have only $n-1$ sources. This differs from the treatment in
\cite{La03} and has important consequences, as we will see in the
next section. We want to evaluate this partition function for
$\hN\rightarrow\infty$, $t=g_s\hN$ fixed, from a saddle point
approximation. We therefore use the path $\tilde\g_{kl}$ from
(\ref{tildegamma}) that was chosen in such a way that the equation
of motion (\ref{eommatrix}) has solutions $s_m^*$ and, for large
$\hN$, $\mathcal{C}_i\subset\g_{p_ip_{i+1}}$. It is only then that
the saddle point expansion converges and makes sense. Obviously
then each integral $\int_{\g}\d\l_m$ splits into a sum
$\sum_{i=1}^{n}\int_{\g_{p_ip_{i+1}}}\d\l_m$. Let
$s^{(i)}\in\mathbb{R}$ be the length coordinate on
$\g_{p_{i}p_{i+1}}$, so that $s^{(i)}$ runs over all of
$\mathbb{R}$. Furthermore, $\tilde\s_i(\l_m)$ only depends on the
number $\hN_i$ of eigenvalues in $\tilde\g_{kl}\cap
D_i=\g_{p_ip_{i+1}}$. Then the partition function
(\ref{partsources}) is a sum of contributions with fixed $\hN_i$
and we rewrite is as
\begin{equation}\label{Z(J)}
Z(g_s,N,J)=\sum_{\stackrel{\sum_i\hN_i=\hN}{\hN_1,\ldots,\hN_{n}}}Z(g_s,\hN,\hN_i)e^{-{1\over
g_s}\sum_{i=1}^{n-1}J_i\hN_i}\ ,
\end{equation}
where now
\begin{eqnarray}
&&\hspace{-0,8cm}Z(g_s,\hN,\hN_i)=\nonumber\\
&&\hspace{-0,8cm}={1\over
\hN_1\ldots\hN_n!}\int_{\g_{p_1p_2}}\d\l^{(1)}_1\ldots
\int_{\g_{p_1p_2}}\d\l^{(1)}_{\hN_1}\ldots\int_{\g_{p_np_{n+1}}}
\d\l^{(n)}_1\ldots\int_{\g_{p_np_{n+1}}}\d\l^{(n)}_{\hN_n}\exp\left(-{t^2\over
g_s^2}S(g_s,t;\l_k^{(i)})\right)\nonumber\\
&&\hspace{-0,8cm}=:\exp\left(\tilde{\mathcal{F}}(g_s,t,\hN_i)\right)
\end{eqnarray}
is the partition function with the additional constraint that
precisely $\hN_i$ eigenvalues lie on $\g_{p_ip_{i+1}}$. Note that
it depends on $g_s,t=g_s\hN$ and $\hN_1,\ldots\hN_{n-1}$ only, as
$\sum_{i=1}^n\hN_i=\hN$. Now that these numbers have been fixed,
there is precisely one solution to the equations of motion, i.e. a
{\it unique} saddle-point configuration, up to permutations of the
eigenvalues, on {\it each} $\g_{p_ip_{i+1}}$. These permutations
just generate a factor $\prod_i \hN_i!$ which cancels the
corresponding factor in front of the integral. As discussed above,
it is important that we have chosen the $\g_{p_ip_{i+1}}$ to
support this saddle point configuration close to the critical
point $\m_i$ of $W$. Moreover, since $\g_{p_ip_{i+1}}$ runs from
one convergence sector to another and by (\ref{condition}) the
saddle point really is dominant (stable), the ``one-loop" and
other higher order contributions are indeed subleading as
$g_s\rightarrow 0$ with $t=g_s\hN$ fixed. This is why we had to be
so careful about the choice of our path $\g$ as being composed of
$n$ pieces $\g_{p_ip_{i+1}}$. In the planar limit
$\n_i:={\hN_i\over\hN}$ is finite, and
$\tilde{\mathcal{F}}(g_s,t,\n_i)={1\over
g_s^2}\tilde{\mathcal{F}}_{0}(t,\n_i)+\ldots$. The saddle point
approximation gives
\begin{equation}\label{F0Seff}
\tilde{\mathcal{F}}_{0}(t,\n_i)= -t^2
S_{eff}(g_s=0,t;s_a^{(j)*}(\n_i))\ ,
\end{equation}
where (cf. (\ref{effaction})) $S_{eff}(g_s=0,t;s_a^{(j)*}(\nu_i))$
is meant to be the value of $S(g_s=0,t;\l(s_a^{(j)*}(\nu_i)))$,
with $\l(s_a^{(j)*}(\nu_i))$ the point on $\g_{p_ip_{i+1}}$
corresponding to the unique saddle point value $s_a^{(j)*}$ with
fixed fraction $\nu_i$ of eigenvalues $\l_m$ in $D_i$. Note that
the ${1\over\hN^2}\sum\dot\l(s_m)$-term in (\ref{effaction})
disappears in the present planar limit. Furthermore, to evaluate
the ``one-loop" term one has to compute the logarithm of the
determinant of an $\hN\times\hN$ matrix which gives a contribution
to $\mathcal{F}(g_s,t,\n_i)$ of order $\hN\sim g_s^{-1}$, as well
as an irrelevant constant $c(\hN)=-{\hN\over2}\log\hN$. The latter
can be absorbed in the overall normalisation of $Z$.

It remains to sum over the $\hN_i$ in (\ref{Z(J)}). In the planar
limit these sums are replaced by integrals:
\begin{eqnarray}
Z(g_s,t,J) &=&\int_0^1\d\n_1\ldots\int_0^1\d
\n_{n}\ \delta\left(\sum_{i=1}^{n}\n_i-1\right)\nonumber\\
&&\exp\left[-{1\over
g_s^2}\left(t\sum_{i=1}^{n-1}J_i\n_i-\tilde{\mathcal{F}}_0(t,\n_i)\right)+
c(\hN)+o(g_s^{-1})\right]\ .
\end{eqnarray}

Once again, in the planar limit, this integral can be evaluated
using the saddle point technique and for the source-dependent free
energy $F(g_s,t,J)={1\over g_s^2} F_0(t,J)+\ldots$ we find
\begin{equation}\label{F0}
F_0(t,J)=\sum_{i=1}^{n-1}J_i\
t\n_i^*-\tilde{\mathcal{F}}_0(t,\n_i^*)\ ,
\end{equation}
where $\n_i^*$ solves the new saddle point equation,
\begin{equation}
tJ_i={\partial\tilde{\mathcal{F}}_0\over\partial\n_i}(t,\n_j)\
.
\end{equation}
This shows that $F_0(t,J)$ is nothing but the Legendre transform
of $\tilde{\mathcal{F}}_0(t,\n_i^*)$ in the $n-1$ latter
variables. If we define
\begin{equation}\label{Snustar}
S_i:=t\n_i^*\ ,\ \mbox{for}\ i=1,\ldots,n-1,
\end{equation}
we have the inverse relation
\begin{equation}\label{SJ}
S_i={\partial F_0\over\partial J_i}(t,J)\ ,
\end{equation}
and with $\mathcal{F}_0(t,S):=\tilde{\mathcal{F}}_0(t,{S_i\over
t})$, where $S:=\{S_1,\ldots,S_{n-1}\}$, one has from (\ref{F0})
\begin{equation}
\mathcal{F}_0(t,S)=\sum_{i=1}^{n-1}J_iS_i-F_0(t,J)\ ,
\end{equation}
where $J_i$ solves (\ref{SJ}). From (\ref{F0Seff}) and the
explicit form of $S_{eff}$, Eq.(\ref{effaction}) with
$\hN\rightarrow\infty$, we deduce that
\begin{equation}\label{mathcalF0}
\mathcal{F}_0(t,S)=t^2\mathcal{P}\int\d s\int\d s'\
\ln(\l(s)-\l(s'))\r_0(s;t,S_i)\r_0(s';t,S_i)-t\int\d s\
W(\l(s))\r_0(s;t,S_i)\ ,
\end{equation}
where $\r_0(s;t,S_i)$ is the eigenvalue density corresponding to
the saddle point configuration $s_a^{(i)*}$ with
${\hN_i\over\hN}=\n_i$ fixed to be $\n_i^*={S_i\over t}$. Hence it
satisfies
\begin{equation}\label{intrhoC}
t\int_{\g^{-1}(\mathcal{C}_i)}\r_0(s;t,S_j)\d s=S_i\ \ \mbox{for}\
i=1,2,\ldots n-1\ ,
\end{equation}
and obviously
\begin{equation}\label{intrhoCn}
t\int_{\g^{-1}(\mathcal{C}_n)}\r_0(s;t,S_j)\d
s=t-\sum_{i=1}^{n-1}S_i\ .
\end{equation}
Note that the integrals in (\ref{mathcalF0}) are convergent and
$\mathcal{F}_0(t,S)$ is a well-defined function.

\subsection{Special Geometry Relations}
After this rather detailed study of the planar limit of
holomorphic matrix models we now turn to the derivation of the
special geometry relations for the Riemann surface
(\ref{Riemannsurface}) and hence the local Calabi-Yau
(\ref{locCY}). Recall that in the matrix model the $S_i=t\n_i^*$
are real and therefore $\mathcal{F}_0(t,S)$ of Eq.
(\ref{mathcalF0}) is a function of {\it real} variables. This is
reflected by the fact that one can generate only a subset of all
possible Riemann surfaces (\ref{Riemannsurface}) from the planar
limit of the holomorphic matrix model, namely those for which
$\n_i^*={1\over4\pi it}\int_{A^i}\zeta$ is real (recall $\zeta=y\d
x$). We are, however, interested in the special geometry of the
most general surface of the form (\ref{Riemannsurface}), which can
no longer be understood as a surface appearing in the planar limit
of a matrix model. Nevertheless, for any such surface we can apply
the formal construction of $\r_0(s)$, which will in general be
complex. Then one can use this complex ``spectral density" to
calculate the function $\mathcal{F}_0(t,S)$ from
(\ref{mathcalF0}), that now depends on {\it complex} variables.
Although this is {\it not} the planar limit of the free energy of
the matrix model, it will turn out to be the prepotential for the
general hyperelliptic Riemann surface (\ref{Riemannsurface}) and
hence of the local Calabi-Yau manifold (\ref{locCY}).

\subsubsection{Rigid special geometry}
Let us then start from the general hyperelliptic Riemann surface
(\ref{Riemannsurface}) which we view as a two-sheeted cover of the
complex plane (cf. Figs.\ref{ABcycles},\ref{albecycles}), with its
 cuts $\mathcal{C}_i$ between $a_i^-$ and $a_i^+$. We choose
a path $\g$ on the upper sheet with parameterisation $\l(s)$ in
such a way that $\mathcal{C}_i\subset\g$. The complex function
$\r_0(s)$ is determined from (\ref{solution}) and
(\ref{y0omega0}), as described above. We define the complex
quantities
\begin{equation}
S_i:={1\over4\pi
i}\int_{A^i}\zeta=t\int_{\g^{-1}(\mathcal{C}_i)}\r_0(s) \ \ \
\mbox{for}\ i=1,\ldots,n-1\ ,
\end{equation}
and the prepotential $\mathcal{F}_0(t,S)$ as in (\ref{mathcalF0})
(of course, $t$ is $-{1\over4}$ times the leading coefficient of
$f_0$ and it can now be complex as well).

Following \cite{La03} one defines the ``principal value of $y_0$"
along the path $\g$ (c.f. (\ref{Pvalue}))
\begin{equation}
y_0^p(s):={1\over2}\lim_{\e\rightarrow0}[y_0(\l(s)+i\e\dot{\l}(s))+y_0(\l(s)-i\e\dot{\l}(s))]\
.
\end{equation}
For points $\l(s)\in\g$ outside $\mathcal{C}:=\cup_i\mathcal{C}_i$
we have $y_0^p(s)=y_0(\l(s))$, while $y_0^p(s)=0$ on
$\mathcal{C}$. With
\begin{equation}
\phi(s):=W(\l(s))-2t\mathcal{P}\int\d
s'\ln(\l(s)-\l(s'))\r_0(s';t,S_i)
\end{equation}
one finds, using (\ref{omegaofrho}), (\ref{y0omega0}) and
(\ref{Pvalue}),
\begin{equation}\label{phiy0}
{d\over ds}\phi(s)=\dot{\l}(s)y_0^p(s)\ .
\end{equation}
The fact that $y_0^p(s)$ vanishes on $\mathcal{C}$ implies
\begin{equation}
\phi(s)=\xi_i:=\mbox{constant on}\ \mathcal{C}_i\ .
\end{equation}
Integrating (\ref{phiy0}) between $\mathcal{C}_i$ and
$\mathcal{C}_{i+1}$ gives
\begin{equation}
\xi_{i+1}-\xi_i=\int_{a^+_i}^{a^-_{i+1}}\d\l \ y_0(\l)={1\over
2}\int_{\b_i}\d x \ y(x)={1\over 2}\int_{\b_i}\zeta\ .
\end{equation}
From (\ref{mathcalF0}) we find for $i<n$
\begin{equation}\label{xinxii}
{\partial\over\partial S_{i}}\mathcal{F}_0(t,S)=-t\int\d
s{\partial\r_0(s;t,S_j)\over\partial S_i}\phi(s)=\xi_n-\xi_i\ .
\end{equation}
To arrive at the last equality we used that $\r_0(s)\equiv0$ on
the complement of the cuts, while on the cuts $\phi(s)$ is
constant and we can use (\ref{intrhoC}) and (\ref{intrhoCn}).
Then, for $i<n-1$,
\begin{equation}\label{F0S}
{\partial\over\partial
S_{i}}\mathcal{F}_0(t,S)-{\partial\over\partial
S_{i+1}}\mathcal{F}_0(t,S)={1\over 2}\int_{\b_i}\zeta\ .
\end{equation}
For $i=n-1$, on the other hand, we find
\begin{equation}
{\partial\over\partial
S_{n-1}}\mathcal{F}_0(t,S)=\xi_{n}-\xi_{n-1}={1\over
2}\int_{\b_{n-1}}\zeta\ .
\end{equation}
We change coordinates to
\begin{equation}
\tilde S_i:=\sum_{j=1}^iS_j\ ,\label{tildeS}
\end{equation}
and find the rigid special geometry relations\footnote{See for
example \cite{CRTP97} for a nice and detailed discussion of the
difference between special and rigid special geometry.}
\begin{eqnarray}
\tilde S_i&=&{1\over 4\pi i }\int_{\a_{i}}\zeta\ \label{sgrela},\\
{\partial\over\partial \tilde S_{i}}\mathcal{F}_0(t,\tilde
S)&=&{1\over2}\int_{\b_{i}}\zeta\ .\label{sgrelb}
\end{eqnarray}
for $i=1,\ldots,n-1$. Note that the basis of one-cycles that
appears in these equations is the one shown in
Fig.\ref{albecycles} and differs from the one used in \cite{La03}.
The origin of this difference is the fact that we introduced only
$n-1$ currents $J_i$ in the partition function
(\ref{partsources}).\\
Next we use the same methods to derive the relation between the
integrals of $\zeta$ over the cycles $\hat \a$ and $\hat \b$ and
the planar free energy.

\subsubsection{Integrals over relative cycles}
The first of these integrals encircles all the cuts, and by
deforming the contour one sees that it is given by the residue of
the pole of $\zeta$ at infinity, which is determined by the
leading coefficient of $f_0(x)$:
\begin{equation}
{1\over4\pi i}\int_{\hat\a}\zeta=t\ .\label{SGhat1}
\end{equation}
The cycle $\hat\beta$ starts at infinity of the lower sheet, runs
to the $n$-th cut and from there to infinity on the upper sheet.
The integral of $\zeta$ along $\hat\b$ is divergent, so we
introduce a (real) cut-off $\L_0$ and instead take $\hat\b$ to run
from $\L_0'$ on the lower sheet through the $n$-th cut to $\L_0$
on the upper sheet. We find
\begin{eqnarray}
{1\over2}\int_{\hat\b}\zeta=\int_{a^+_{n}}^{\L_0}
y_0(\l)\d\l&=&\phi(\l^{-1}(\L_0))-\phi(\l^{-1}(a^+_{n}))=\phi(\l^{-1}(\L_0))-\xi_{n}\nonumber\\
&=&W(\L_0)-2t\mathcal{P}\int\d s'\ln(\L_0-\l(s'))\r_0(s';t,\tilde
S_i)-\xi_{n}.
\end{eqnarray}
On the other hand we can calculate
\begin{equation}\label{xin}
{\partial\over\partial t}\mathcal{F}_0(t,\tilde S)=-\int\d s\
\phi(\l(s)){\partial\over\partial t}[t\r_0(s;t,\tilde
S_i)]=-\sum_{i=1}^n\xi_i\int_{\g^{-1}(\mathcal{C}_i)}\d s\
{\partial\over\partial t}[t\r_0(s;t,\tilde S_i)]=-\xi_{n}\ ,
\end{equation}
(where we used (\ref{intrhoC}) and (\ref{intrhoCn})) which leads
to
\begin{eqnarray}
{1\over2}\int_{\hat \b}\zeta&=&{\partial\over\partial
t}\mathcal{F}_0(t,\tilde
S)+W(\L_0)-2t\mathcal{P}\int\d s'\ln(\L_0-\l(s'))\r_0(s';t,\tilde S_i)\nonumber\\
&=&{\partial\over\partial t}\mathcal{F}_0(t,\tilde
S)+W(\L_0)-t\log\L_0^2+o\left({1\over\L_0}\right)\ .\label{SGhat2}
\end{eqnarray}
Together with (\ref{SGhat1}) this looks very similar to the usual
special geometry relation. In fact, the cut-off independent term
is the one one would expect from special geometry. However, the
equation is corrected by cut-off dependent terms. The last terms
vanishes if we take $\L_0$ to infinity but there remain two
divergent terms which we want to interpret in section
\ref{superpot}.\footnote{Of course, one could define a cut-off
dependent function $\mathcal{F}^{\L_0}(t,\tilde
S):=\mathcal{F}_0(t,\tilde S)+tW(\L_0)-{t^2\over2}\log\L_0^2$ for
which one has
${1\over2}\int_{\hat\b}\zeta={\partial\mathcal{F}^{\L_0}\over\partial
t}+o\left({1\over\L_0}\right)$ similar to \cite{DOV04}. Note,
however, that this is not a standard special geometry relation due
to the presence of the $o\left({1\over\L_0}\right)$-terms.
Furthermore $\mathcal{F}^{\L_0}$ has no interpretation in the
matrix model and is divergent as $\L_0\rightarrow\infty$.}

\subsubsection{Homogeneity of the prepotential}
The prepotential on the moduli space of complex structures of a
{\it compact} Calabi-Yau manifold is a holomorphic function that
is homogeneous of degree two. On the other hand, the structure of
the local Calabi-Yau manifold (\ref{locCY}) is captured by a
Riemann surface and it is well-known that these are related to
rigid special geometry. The prepotential of rigid special
manifolds does not have to be homogeneous (see for example
\cite{CRTP97}), and it is therefore important to explore the
homogeneity structure of $\mathcal{F}_0(t,\tilde S)$. The result
is quite interesting and it can be written in the form
\begin{equation}\label{homogeneity}
\sum_{i=1}^{n-1}\tilde S_i{\partial\mathcal{F}_0\over\partial
\tilde{S_i}}(t,\tilde S_i)+t{\partial\mathcal{F}_0\over\partial
t}(t,\tilde S_i)=2\mathcal{F}_0(t,\tilde S_i)+t\int\d s\
\r_0(s;t,\tilde S_i)W(\l(s))\ .
\end{equation}
To derive this relation we rewrite Eq.(\ref{mathcalF0}) as
\begin{eqnarray}
2\mathcal{F}_0(t,\tilde S_i)&=&-t\int\d s\ \r_0(s;t,\tilde
S_i)\left[\phi(s)+W(\l(s))\right]\nonumber\\
&=&-t\int\d s\ \r_0(s;t,\tilde
S_i)W(\l(s))+\sum_{i=1}^{n-1}(\xi_n-\xi_i)S_i-t\xi_n\ .
\end{eqnarray}
Furthermore, we have $\sum_{i=1}^{n-1}\tilde
S_i{\partial\mathcal{F}_0\over\partial\tilde S_i}(t,\tilde
S_i)=\sum_{i=1}^{n-1} S_i{\partial\mathcal{F}_0\over\partial
S_i}(t, S_i)=\sum_{i=1}^{n-1}S_i(\xi_n-\xi_i)$, where we used
(\ref{xinxii}). The result then follows from (\ref{xin}).\\
Of course, the prepotential was not expected to be homogeneous,
since already for the simplest example, the conifold,
$\mathcal{F}_0$ is known to be non-homogeneous (see section
\ref{conifold}). However, Eq.(\ref{homogeneity}) shows the precise
way in which the homogeneity relation is modified on the local
Calabi-Yau manifold (\ref{locCY}).

\subsubsection{Duality transformations}
The choice of the basis $\{\a^i,\b_j,\hat\a,\hat\b\}$ for the
(relative) one-cycles on the Riemann surface was particularly
useful in the sense that the integrals over the compact cycles
$\a^i$ and $\b_j$ reproduce the familiar rigid special geometry
relations, whereas new features appear only in the integrals over
$\hat\a$ and $\hat\b$. In particular, we may perform any
symplectic transformation of the compact cycles $\a^i$ and $\b_j$,
$i,j=1,\ldots n-1$, among themselves to obtain a new set of
compact cycles which we call $a^i$ and $b_j$. Such symplectic
transformations can be generated from (i) transformations that do
not mix $a$-type and $b$-type cycles, (ii) transformations
$a^i=\a^i,\ b_i=\b_i+\a^i$ for some $i$ and (iii) transformations
$a^i=\b_i$, $b_i=-\a^i$ for some $i$. (These are analogue to the
trivial, the $T$ and the $S$ modular transformations of a torus.)
For transformations of the first type the prepotential
$\mathcal{F}$ remains unchanged, except that it has to be
expressed in terms of the new variables $s_i$, which are the
integrals of $\zeta$ over the new $a^i$ cycles. Since the
transformation is symplectic, the integrals over the new $b_j$
cycles then automatically are the derivatives
${\partial\mathcal{F}_0(t,s)\over\partial s_i}$. For
transformations of the second type the new prepotential is given
by $\mathcal{F}_0(t,\tilde S_i)+i\pi\tilde S_i^2$ and for
transformations of the third type the prepotential is a Legendre
transform with respect to $\int_{a^i}\zeta$. In the corresponding
gauge theory the latter transformations realise electric-magnetic
duality. Consider e.g. a symplectic transformation that exchanges
all compact $\a^i$-cycles with all compact $\b_i$ cycles:
\begin{equation}
\left(\begin{array}{c}\a^i\\\b_i\end{array}\right)\rightarrow\left(\begin{array}{c}a^i\\
b_i\end{array}\right)=\left(\begin{array}{c}\b_i\\-\a^i\end{array}\right),
\ \ \ i=1,\ldots n-1 \ .
\end{equation}
Then the new variables are the integrals over the $a_i$-cycles
which are
\begin{equation}
\tilde\pi_i:={1\over2}\int_{\b_i}\zeta={\partial\mathcal{F}_0(t,\tilde
S)\over\partial \tilde S_i}
\end{equation}
and the dual prepotential is given by the Legendre transformation
\begin{equation}
\mathcal{F}_{D}(t,\tilde\pi):=\sum_{i=1}^{n-1}\tilde
S_i\tilde\pi_i-\mathcal{F}_0(t,\tilde S)\ ,
\end{equation}
such that the new special geometry relation is
\begin{equation}
{\partial\mathcal{F}_D(t,\tilde\pi)\over\partial\tilde\pi_i}=\tilde
S_i={1\over4\pi i}\int_{\a^i}\zeta\ .
\end{equation}
Comparing with (\ref{F0}) one finds that
$\mathcal{F}_D(t,\tilde\pi)$ actually coincides with $F_0(t,J)$
where $J_i-J_{i+1}=\tilde\pi_i$ for $i=1,\ldots n-2$ and
$J_{n-1}=\tilde\pi_{n-1}$.

Next, let us see what happens if we also include symplectic
transformations involving the relative cycles $\hat\a$ and
$\hat\b$. An example of a transformation of type (i) that does not
mix $\{\a,\hat\a\}$ with $\{\b,\hat\b\}$ cycles is the one from
$\{\a^i,\b_j,\hat\a,\hat\b\}$ to $\{A^i,B_j\}$, c.f.
Figs.\ref{ABcycles},\ref{albecycles}. This corresponds to
\begin{eqnarray}
\bar S_1&:=&\tilde S_1\ ,\nonumber\\
\bar S_i&:=&\tilde S_i-\tilde S_{i-1}\ \ \ \mbox{for}\ \ i=2,\ldots n-1\ ,\label{barS}\\
\bar S_n&:=&t-\tilde S_{n-1}\ ,\nonumber
\end{eqnarray}
so that
\begin{equation}
\bar S_i={1\over 4\pi i}\int_{A^i}\zeta\ .
\end{equation}
The prepotential does not change, except that it has to be
expressed in terms of the $\bar S_i$. One then finds for
$B_i=\sum_{j=i}^{n-1}\b_j+\hat\b$
\begin{equation}
{1\over2}\int_{B_i}\zeta={\partial\mathcal{F}_0(t,\bar
S)\over\partial \bar S_i}+W(\L_0)-\left(\sum_{j=1}^n\bar
S_i\right)\log\L_0^2+o\left({1\over\L_0}\right)\ .
\end{equation}
We see that as soon as one ``mixes" the cycle $\hat\b$ into the
set $\{\b_i\}$ one obtains a number of relative cycles $B_i$ for
which the special geometry relations are corrected by cut-off
dependent terms. An example of transformation of type (iii) is
$\hat\a\rightarrow\hat\b$, $\hat\b\rightarrow-\hat\a$. Instead of
$t$ one then uses
\begin{equation}
\hat\pi:={\partial\mathcal{F}_0(t,\tilde S)\over\partial
t}=\lim_{\L_0\rightarrow\infty}\left[{1\over2}\int_{\hat\b}\zeta-W(\L_0)+t\log\L_0^2\right]
\end{equation}
as independent variable and the Legendre transformed prepotential
is
\begin{equation}
\hat{\mathcal{F}}(\hat\pi,\tilde
S):=t\hat\pi-\mathcal{F}_0(t,\tilde S)\ ,
\end{equation}
so that now
\begin{equation}
{\partial\hat{\mathcal{F}}(\hat\pi,\tilde
S)\over\partial\hat\pi}=t={1\over4\pi i}\int_{\hat\a}\zeta\ .
\end{equation}
Note that the prepotential is well-defined and independent of the
cut-off in all cases (in contrast to the treatment in
\cite{DOV04}). The finiteness of $\hat{\mathcal{F}}$ is due to
$\hat\pi$ being the {\it corrected, finite} integral over the
relative $\hat\b$-cycle.\\
Note also that if one exchanges all coordinates simultaneously,
i.e. $\a^i\rightarrow\b_i,\
\hat\a\rightarrow\hat\b,\b_i\rightarrow-\a^i,\
\hat\b\rightarrow-\hat\a$, one has
\begin{equation}
\hat{\mathcal{F}}_D(\hat\pi,\tilde\pi):=t\hat\pi+\sum_{i=1}^{n-1}\tilde
S_i\tilde\pi_i-\mathcal{F}_0(t,\tilde S_i)\ .
\end{equation}
Using the generalised homogeneity relation (\ref{homogeneity})
this can be rewritten as
\begin{equation}
\hat{\mathcal{F}}_D(\hat\pi,\tilde\pi)=\mathcal{F}_0(t,\tilde
S_i)+t\int\d s\ \r_0(s;t,\tilde S_i)W(\l(s))\ .
\end{equation}

It would be quite interesting to understand the results of this
chapter concerning the parameter spaces of local Calabi-Yau
manifolds in a more geometrical way in the context of (rigid)
special K\"ahler manifolds along the lines of \cite{CRTP97}.

\section{The superpotential}\label{superpot}
\setcounter{equation}{0} Adding a background three-form flux to
type IIB strings on a local Calabi-Yau manifold generates an
effective superpotential and breaks the $\mathcal{N}=2$
supersymmetry of the effective action to $\mathcal{N}=1$. Starting
from the usual formula for the effective superpotential
\cite{CIV01} and performing a change of basis one arrives at
\begin{equation}\label{Weff}
W_{eff}=\sum_{i=1}^{n-1}\left(\int_{\G_{\a^i}}H\int_{\G_{\b_i}}\O-\int_{\G_{\b_i}}
H\int_{\G_{\a^i}}\O\right)+\left(\int_{\G_{\hat\a}}H\int_{\G_{\hat\b}}\O-\int_{\G_{\hat\b}}
H\int_{\G_{\hat\a}}\O\right)\ .
\end{equation}
As explained before, the integrals over three-cycles reduce to
integrals over the one-cycles in $H_1(\S,\{Q,Q'\})$ on the Riemann
surface $\S$. But this implies that the divergent terms in
(\ref{SGhat2}) are quite problematic, as they lead to a divergence
of the superpotential which has to be removed for the potential to
make sense.

\subsection{Pairings on Riemann surfaces with marked points}
To understand the divergence somewhat better we will study the
meromorphic one-form $\zeta$ in more detail. First of all we
observe that the integrals $\int_{\a^i}\zeta$ and
$\int_{\b_j}\zeta$ only depend on the cohomology class $[\zeta]\in
H^1(\hat\S)$, whereas $\int_{\hat\b}\zeta$  (where $\hat\b$
extends between the poles of $\zeta$, i.e. from $\infty'$ on the
lower sheet, corresponding to $Q'$, to $\infty$ on the upper
sheet, corresponding to $Q$,) is not only divergent, it also
depends on the representative of the cohomology class, since for
$\tilde\zeta= \zeta+\d\rho$ one has
$\int_{\hat\b}\tilde\zeta=\int_{\hat\b}\zeta+
\int_{\partial\hat\b}\r\left(\neq\int_{\hat\b}\zeta\right)$. Note
that the integral would be independent of the choice of the
representative if we constrained $\r$ to be zero at
$\partial\hat\b$. But as we marked $Q,Q'$ on the Riemann surface,
$\r$ is allowed to take finite or even infinite values at these
points and therefore the integrals differ in general.

The origin of this complication is, of course, that our cycles are
elements of the {\it relative} homology group $H_1(\S,\{Q,Q'\})$.
Then, their is a natural map
$\langle.,.\rangle:H_1(\S,\{Q,Q'\})\times
H^1(\S,\{Q,Q'\})\rightarrow \mathbb{C}$. $H^1(\S,\{Q,Q'\})$ is the
relative cohomology group dual to $H_1(\S,\{Q,Q'\})$. In general,
on a manifold $M$ with submanifold $N$, elements of relative
cohomology can be defined as follows (see for example
\cite{KL87}). Let $\O^k(M,N)$ be the set of $k$-forms on $M$ that
vanish on $N$. Then $H^k(M,N):=Z^k(M,N)/B^k(M,N)$, where
$Z^k(M,N):=\{\o\in\O^k(M,N):\d\o=0\}$ and
$B^k(M,N):=\d\O^{k-1}(M,N)$. For $[\hat\G]\in H_k(M,N)$ and
$[\eta]\in H^k(M,N)$ the pairing is defined as
\begin{equation}
\langle\hat\G,\eta\rangle:=\int_{\hat \G}\eta\ .
\end{equation}
This does not depend on the representative of the classes, since
the forms are constraint to vanish on $N$.\\
Now consider $\xi\in\O^k(M)$ such that $i^*\xi=\d\phi$, where $i:N\rightarrow M$
is the inclusion mapping. Note that $\xi$ is not
a representative of an element of relative cohomology, as it does not vanish on
$N$. However, there is another representative in its cohomology class
$[\xi]\in H^k(M)$, namely $\xi_{\phi}=\xi-\d\phi$ which now is also a
representative of
$H^k(M,N)$. For elements $\xi$ with this property we can extend the definition
of the pairing to
\begin{equation}
\langle \hat\G,\xi\rangle:=\int_{\hat \G}\left(\xi-\d\phi\right)\
.
\end{equation}

Clearly, the one-form $\zeta=y\d x$ on $\hat\S$ is not a
representative of an element of $H^1(\S,\{Q,Q'\})$. According to
the previous discussion, one might try to find
$\zeta_{\varphi}=\zeta-\d\varphi$ where $\varphi$ is chosen in
such a way that $\zeta_{\varphi}$ vanishes at $Q,Q'$, so that in
particular $\int_{\hat\b}\zeta_{\varphi}=\mbox{finite}$. In other
words we would like to find a representative of $[\zeta]\in
H^1(\hat\S)$ which is also a representative of $H^1(\S,\{Q,Q'\})$.
Unfortunately, this is not possible, because of the logarithmic
divergence, i.e. the simple poles at $Q,Q'$, which cannot be
removed by an exact form. The next best thing we can do instead is
to determine $\varphi$ by the requirement that
$\zeta_{\varphi}=\zeta-\d\varphi$ only has simple poles at $Q,Q'$.
Then we define the pairing
\begin{equation}\label{pair}
\left\langle\hat\b,\zeta\right\rangle:=\int_{\hat\b}\left(\zeta-\d\varphi\right)=
\int_{\hat\b}\zeta_{\varphi}\ ,
\end{equation}
which diverges only logarithmically. To regulate this divergence
we introduce a cut-off as before, i.e. we take $\hat\b$ to run
from $\L_0'$ to $\L_0$. We will have more to say about this
logarithmic divergence in the next section. So although
$\zeta_{\varphi}$ is not a representative of a class in
$H^1(\S,\{Q,Q'\})$ it is as close as we can get.

We now want to determine $\varphi$ explicitly. To keep track of
the poles and zeros of the various terms it is useful to apply the
theory of divisors, as explained e.g. in \cite{FK}. Let
$P_1,\ldots P_{2\gh+2}$ denote those points on the Riemann surface
of genus $\gh=n-1$ that correspond to the zeros of $y$ (i.e. to
the $a_i^\pm$). Close to $a_i^\pm$ the good coordinates are
$z_i^\pm=\sqrt{x-a_i^\pm}$. This shows that the divisor of $y$ is
${P_1\dots P_{2\gh+2}\over Q^{\gh+1}Q'^{\gh+1}}$, which simply
states that $y$ has simple zeros at $P_1,\ldots P_{2\gh+2}$ and
poles of order up to $\gh+1$ at $Q$ and $Q'$. Let $R,R'$ be the
points on the Riemann surface that correspond to zero on the upper
and lower sheet, respectively, then the divisor of $x$ is given by
${RR'\over QQ'}$. Finally the divisor of $\d x$ is ${P_1\dots
P_{2\gh+2}\over Q^2Q'^2}$, since close to $a_i^\pm$ one has $\d
x\sim z_i^\pm\d z_i^\pm$, and obviously $\d x$ has double poles at
$Q$ and $Q'$. To study the leading poles and zeros of combinations
of these quantities one simply has to multiply the corresponding
divisors. In particular, the divisor of ${\d x\over y}$ is
$Q^{\gh-1}Q'^{\gh-1}$ and for $\gh\geq 1$ it has no poles. Also,
the divisor of $\zeta=y^2{\d x\over y}$ is ${P_1^2\ldots
P_{2\gh+2}^2\over Q^{\gh+3}Q'^{\gh+3}}$ showing that $\zeta$ has
poles of order
$\gh+3,\gh+2,\ldots 2,1$ at $Q$ and $Q'$.\\
Consider now $\varphi_k:={x^k\over y}$ with $\d\varphi_k={{k
x^{k-1}\d x}\over y}-{x^k {y^2}'\d x\over2 y^3}$. For $x$ close to
$Q$ or $Q'$ the leading term of this expression is
$\pm(k-\gh-1)x^{k-\gh-2}\d x.$ This has no pole at $Q,Q'$ for
$k\leq \gh$, and for $k=\gh+1$ the coefficient vanishes, so that
we do not get simple poles at $Q,Q'$. This is as expected as
$\d\varphi_k$ is exact and cannot contain poles of first order.
For $k\geq\gh+2=n+1$ the leading term has a pole of order $k-\gh$
and so $\d\varphi_k$ contains poles of order
$k-\gh,k-\gh-1,\ldots2$ at $Q,Q'$. Note also that at $P_1,\ldots
P_{2\gh+2}$ one has double poles for all $k$ (unless a zero of $y$
occurs at $x=0$). Next, we set
\begin{equation}
\varphi={\mathcal{P}\over y} ,
\end{equation}
with $\mathcal{P} $ a polynomial of order $2\gh+3$. Then $\d
\varphi$ has poles of order $\gh+3,\gh+2,\ldots 2$ at $Q,Q'$, and
double poles at the zeros of $y$ (unless a zero of $\mathcal{P}_k$
coincides with one of the zeros of $y$). From the previous
discussion it is clear that we can choose the coefficients in
$\mathcal{P}$ such that $\zeta_{\varphi}=\zeta-\d\varphi$ only has
a simple pole at $Q,Q'$ and double poles at $P_1,P_2,\ldots
P_{2\gh+2}$. Actually, the coefficients of the monomials $x^k$ in
$\mathcal{P}$ with $k\leq\gh$ are not fixed by this requirement.
Only the $\gh+2$ highest coefficients will be determined, in
agreement with the fact that we cancel the $\gh+2$ poles of order
$\gh+3,\ldots2$.

It remains to determine the polynomial $\mathcal{P}$ explicitly. The part of $\zeta$ contributing to the poles of order $\geq2$ at $Q,Q'$ is easily seen to be $\pm W'(x)\d x$ and we obtain the condition
\begin{equation}
W'(x)-\left(\mathcal{P}(x)\over\sqrt{W'(x)^2+f(x)}\right)'=
o\left(1\over x^2\right)\ .
\end{equation}
Integrating this equation, multiplying by the square root and developing the square root leads to
\begin{equation}
W(x)W'(x)-{2t\over {n+1}}x^{n}-\mathcal{P}(x)=cx^{n}+
o\left(x^{n-1}\right)\ ,
\end{equation}
where $c$ is an integration constant.
We read off
\begin{equation}\label{varphi}
\varphi(x)={W(x)W'(x)-\left({2t\over {n+1}}+c\right)x^{n}+
o\left(x^{n-1}\right)\over y}\ ,
\end{equation}
and in particular, for $x$ close to infinity on the upper or lower sheet,
\begin{equation}
\varphi(x)\sim\pm\left[W(x)-c+
o\left(1\over x\right)\right]\ .\label{philarge}
\end{equation}
The arbitrariness in the choice of $c$ has to do with the fact
that the constant $W(0)$ does not appear in the description of the
Riemann surface. In the sequel we will choose $c=0$, such that the
full $W(x)$ appears in (\ref{philarge}). As is clear from our
construction, and is easily verified explicitly, close to $Q,Q'$
one has $\zeta_{\varphi}\sim\left(\mp{2t\over x}+o\left({1\over
x^2}\right)\right)\d x$.

With this $\varphi$ we find
\begin{equation}\label{intzetaphi}
\int_{\hat\b}\zeta_{\varphi}=\int_{\hat\b}\zeta-\int_{\hat\b}\d
\varphi=
\int_{\hat\b}\zeta-\varphi(\L_0)+\varphi(\L_0')=\int_{\hat\b}\zeta-2\left(W(\L_0)+o\left(1\over\L_0\right)\right)\
.
\end{equation}
Note that, contrary to $\zeta$, $\zeta_{\varphi}$ has poles at the
zeros of $y$, but these are double poles and it does not matter
how the cycle is chosen with respect to the location of these
poles (as long as it does not go right through the poles). Note
also that we do not need to evaluate the integral of
$\zeta_{\varphi}$ explicitly. Rather one can use the known result
(\ref{SGhat2}) for the integral of $\zeta$ to find from
(\ref{intzetaphi})
\begin{equation}\label{resultpairing}
{1\over2}\left\langle\hat\b,\zeta\right\rangle={1\over2}\int_{\hat\b}\zeta_{\varphi}={\partial\over\partial
t}\mathcal{F}_0(t,\tilde
S)-t\log\L_0^2+o\left({1\over\L_0}\right)\ .
\end{equation}

Finally, let us comment on the independence of the representative
of the class $[\zeta]\in H^1(\hat\S)$. Suppose we had started from
$\tilde\zeta:=\zeta +\d\rho$ instead of $\zeta$. Then determining
$\tilde\varphi$ by the same requirement that
$\tilde\zeta-\d\tilde\varphi$ only has first order poles at $Q$
and $Q'$ would have led to $\tilde\varphi=\varphi+\rho$ (a
possible ambiguity related to the integration constant $c$ again
has to be fixed). Then obviously
\begin{equation}
\left\langle\hat\b,\tilde\zeta\right
\rangle=\int_{\hat\b}\tilde\zeta-\int_{\partial\hat\b}\tilde\varphi=\int_{\hat\b}\zeta-\int_{\partial\hat\b}\varphi=\left\langle\hat\b,\zeta\right\rangle\
,
\end{equation}
and hence our pairing only depends on the cohomology class
$[\zeta]$.

\subsection{The superpotential revisited}
At last we turn to the effective superpotential $W_{eff}$ of the
low energy gauge theory given by the integrals of the three-forms
$\O$ and $H$ over the three-cycles of the Calabi-Yau manifold
(c.f. Eq.(\ref{Weff})). Following \cite{CIV01} and \cite{DV02} we
define for the integrals of $H$ over the cycles $\G_A$ and $\G_B$:
\begin{equation}
N_i=\int_{\G_{A^i}}H\ \ ,\ \ \tau_i=
\left\langle\G_{B_i},H\right\rangle\ \ \ \ \mbox{for}\ \
i=1,\ldots n\ .
\end{equation}
It follows for the integrals over the cycles $\G_{\a}$ and $\G_{\b}$
\begin{eqnarray}
\tilde N_i=\int_{\G_{\a^i}}H=\sum_{j=1}^iN_j\ \ &,&\ \ \tilde
\tau_i=\int_{\G_{\b_i}}H=\tau_i-\tau_{i+1}\ \ \ \ \mbox{for}\ \
i=1,\ldots n-1\
,\nonumber\\
N=\sum_{i=1}^nN_i= \int_{\G_{\hat\a}}H\ \ &,&\ \
\tilde\tau_0=\left\langle\G_{\hat\b},H\right\rangle=\tau_n\ .
\end{eqnarray}
For the non-compact cycles, instead of the usual integrals, we use
the pairings of the previous section. On the Calabi-Yau, the
pairings are to be understood e.g. as $\tau_i=-i\pi\left\langle
B_i,h\right\rangle$, where $\int_{S^2}H=-2\pi i h$ and $S^2$ is
the sphere in the fibre of $\G_{B_i}\rightarrow B_i$. Note that
this implies that the $\tau_i$ as well as $\tilde\tau_0$ have (at
most) a logarithmic divergence, whereas the $\tilde\tau_i$ are
finite. We propose that the superpotential should be defined as
\begin{eqnarray}\label{superpotential}
W_{eff}&=&\sum_{i=1}^{n-1}\left(\int_{\G_{\a^i}}H\int_{\G_{\b_i}}\O-\int_{\G_{\b_i}}
H\int_{\G_{\a^i}}\O\right)+\left(\int_{\G_{\hat\a}}H\cdot\left\langle\G_{\hat\b},
\O\right\rangle-\left\langle\G_{\hat\b},H\right\rangle\int_{\G_{\hat\a}}\O\right)\nonumber\\
&=&-i\pi \sum_{i=1}^{n-1}\left(\tilde N_i\int_{\b_i}\zeta-
\tilde\tau_i\int_{\a^i}\zeta\right)-i\pi \left(N
\left\langle\hat\b,\zeta\right\rangle-\tilde\tau_0\int_{\hat\a}\zeta\right)\
.
\end{eqnarray}
This formula is very similar to the one advocated for example in
\cite{LMW02}, but now the pairing (\ref{pair}) is to be used. Note
that Eq.(\ref{superpotential}) is invariant under symplectic
transformations on the basis of (relative) three-cycles on the
local Calabi-Yau manifold, resp. (relative) one-cycles on the
Riemann surface, provided one uses the pairing (\ref{pair}) for
every relative cycle. These include $\a^i\rightarrow\b_i,\
\hat\a\rightarrow\hat\b,\ \b_i\rightarrow-\a^i,\
\hat\b\rightarrow-\hat\a$, which acts as electric-magnetic
duality. Using the special geometry relations (\ref{sgrela}),
(\ref{sgrelb}) for the standard cycles and (\ref{SGhat1}),
(\ref{resultpairing}) for the relative cycles, we obtain
\begin{eqnarray}
-{1\over2\pi i}W_{eff}&=&\sum_{i=1}^{n-1}\tilde
N_i{\partial\over\partial \tilde S_i}\mathcal{F}_0(t,\tilde
S_1,\ldots,\tilde S_{n-1})-2\pi
i\sum_{i=1}^{n-1}\tilde\tau_i\tilde
S_i\nonumber\\&&+N{\partial\over\partial t}\mathcal{F}_0(t,\tilde
S_1,\ldots,\tilde S_{n-1})-\left(N\log \L^2_0+2\pi
i\tilde\tau_0\right)t+o\left({1\over\L_0}\right) \
.\nonumber\\\label{WL0}
\end{eqnarray}
The limit $\L_0\rightarrow \infty$ can now be taken provided
\begin{equation}
N\log\L_0^2+2\pi i\tilde\tau_0=N\log\L^2+2\pi i\tau\label{renorm}
\end{equation}
with finite $\L$ and $\tau$. Indeed, $\tilde\tau_0$ is the only
flux number in (\ref{WL0}) that depends on $\L_0$, and its
divergence is logarithmic because of its definition as a pairing
$\left\langle\hat\b,h\right\rangle$. It is, of course, its
interpretation as the $SU(N)$ gauge coupling constant
$\tilde\tau_0={\t_0\over 2\pi}+{4\pi i\over g_0^2}$ which ensures
the exact cancellation of the $\log\L_0$ -terms.

Eq.(\ref{WL0}) can be brought into the form of \cite{DV02} if we
use the coordinates $\bar S_i$, as defined in (\ref{barS}) and
such that $\bar S_i={1\over4\pi i}\int_{A^i}\zeta$ for all
$i=1,\ldots n$. We get
\begin{eqnarray}
-{1\over2\pi i}W_{eff}&=&\sum_{i=1}^{n} N_i{\partial\over\partial
\bar
S_i}\mathcal{F}_0(\bar S)\nonumber\\
&&-\sum_{i=1}^{n-1}\bar S_i\left(2\pi
i\sum_{j=i}^{n-1}\tilde\tau_j+N\log\L^2+2\pi i\tau\right)-\bar
S_{n}(2\pi i\tau+N\log\L^2)\ .\nonumber\\
\end{eqnarray}
Setting
\begin{equation}
2\pi i\sum_{j=i}^{n-1}\tilde\tau_j+N\log\L^2=N_i\log\L_i^2\ \ \
\mbox{for}\ i\in\{1,\ldots n-1\}
\end{equation}
and $\L_n:=\L$ we arrive at
\begin{equation}\label{DVsuperpotential}
-{1\over2\pi i}W_{eff}=\sum_{i=1}^{n}\left[
N_i{\partial\over\partial \bar S_i}\mathcal{F}_0(\bar S)-\bar
S_i\log\L_i^{2N_i} -2\pi i\bar S_i\tau\right]\ .
\end{equation}
This coincides with the corresponding formula in \cite{DV02}
provided we identify ${\partial\mathcal{F}_0(S)\over\partial \bar
S_i}$ with ${\partial\mathcal{F}_0^{pert}(S)\over\partial \bar
S_i}+\bar S_i\log\bar S_i$. Indeed, $\mathcal{F}_0^{pert}$ was the
perturbative part of the free energy of the matrix model and it
was argued in \cite{DV02} that the $S\log S$ term comes from the
measure. Here instead, $\mathcal{F}_0$ is computed in the exact
planar limit of the matrix model, including perturbative and
non-perturbative terms and therefore the $S_i\log S_i$-terms are
already included.\footnote{The presence of $\bar S_i\log \bar S_i$
in ${\partial\mathcal{F}_0\over\partial\bar S_i}$ and hence in
$\int_{B_i}\zeta$ can be easily proven by monodromy arguments
\cite{CIV01}.}

Finally, note that we could have chosen $\hat\b$ to run from a
point $\L_0'=|\L_0|e^{i\t/2}$ on the lower sheet to a point
$\L_0=|\L_0|e^{i\t/2}$ on the upper sheet. Then one would have
obtained an additional term $-it\t$, on the right-hand side of
(\ref{resultpairing}), which would have led to
\begin{equation}
\tau\rightarrow\tau+N{\t\over2\pi}
\end{equation}
in (\ref{DVsuperpotential}), as expected.

\subsection{Example: the conifold}\label{conifold}
Next we want to illustrate our general discussion by looking at
the simplest example, i.e. $n=1$. If we take $W={x^2\over2}$ and
$f(x)=-\m=-4t$, $\mu\in\mathbb{R}^+$, the local Calabi-Yau is
nothing but the deformed conifold,
\begin{equation}
x^2+v^2+w^2+z^2-\m=0\ .
\end{equation}
As $n=1$ the corresponding Riemann surface has genus zero. Then
\begin{equation}
\zeta=y\d x=\left\{\begin{array}{c}\ \ \sqrt{x^2-4t}\ \d x\ \ \
\mbox{on the upper sheet}\\-\sqrt{x^2-4t}\ \d x\ \ \ \mbox{on the
lower sheet}\end{array}\right.\ .
\end{equation}
We have a cut $\mathcal{C}=[-2\sqrt{t},2\sqrt{t}]$ and take
$\l(s)=s$ to run along the real axis. The corresponding $\r_0(s)$
is immediately obtained from (\ref{solution}) and (\ref{y0omega0})
and yields the well-known $\r_0(s)={1\over2\pi t}\sqrt{4t-s^2}$,
for $s\in[-2\sqrt{t},2\sqrt{t}]$ and zero otherwise, and from
(\ref{mathcalF0}) we find the planar free energy
\begin{equation}
\mathcal{F}_0(t)={t^2\over2}\log t-{3\over4}t^2\
.\label{F0conifold}
\end{equation}
Note that $t\int\d s\,\r_0(s)W(\l(s))={t^2\over2}$ and
$\mathcal{F}_0$ satisfies the generalised homogeneity relation
(\ref{homogeneity})
\begin{equation}
t{\partial\mathcal{F}_0\over\partial
t}(t)=2\mathcal{F}_0(t)+{t^2\over2}\ .
\end{equation}

Obviously one has $\zeta=-2t{\d x\over
y}+\d\left(xy\over2\right)$, which would correspond to
$\varphi={xy\over2}$. Comparing with (\ref{varphi}) this would
yield $c=t$. The choice $c=0$ instead leads to
$\varphi={xy\over2}+t{x\over y}$ and $\zeta=-2t{\d x\over
y}+4t^2{\d x\over y^3} +\d\varphi$. The first term has a pole at
infinity and leads to the logarithmic divergence, while the second
term has no pole at infinity but second order poles at
$\pm2\sqrt{t}$. One has
\begin{eqnarray}
\int_{\hat\a}\zeta&=&4\pi i t=4\pi i\bar S\\
\int_{\hat\b}\zeta&=&\L_0\sqrt{\L_0^2-4t}-4t\log\left({\L_0\over2\sqrt t}+\sqrt{{\L_0^2\over4t}-1}\right)\\
2\varphi(\L_0)&=&\L_0\sqrt{\L_0^2-4t}+2t{\L_0\over\sqrt{\L_0^2-4t}}\
.
\end{eqnarray}
Then
\begin{eqnarray}
{1\over2}\left\langle\hat\b,\zeta\right\rangle
&=&t\log\left({4t\over\L_0^2}\right)-2t\log\left(1+\sqrt{{1-{4t\over\L_0^2}}}\right)-t{1\over\sqrt{1-{4t\over\L_0^2}}}\nonumber\\
&=&{\partial\mathcal{F}_0(t)\over \partial
t}-t\log\L_0^2+o\left({1\over\L_0^2}\right)\ ,
\end{eqnarray}
where we used the explicit form of $\mathcal{F}_0(t)$,
(\ref{F0conifold}). Finally, in the present case,
Eq.(\ref{superpotential}) for the superpotential only contains the
relative cycles,
\begin{equation}
W_{eff}=-i\pi
N\left\langle{\hat\b},\zeta\right\rangle+i\pi\tilde\tau_0\int_{\hat\a}\zeta
\end{equation}
or ($\bar S=t$)
\begin{eqnarray}
-{1\over2\pi i}W_{eff}&=&N\left(t\log t-t-t\log\L_0^2\right)-2\pi i\tilde\tau_0t+o\left({1\over\L_0^2}\right)\nonumber\\
&=&\bar S\log\left({\bar S^N\over\L^{2N}}\right)-\bar SN -2\pi
i\tau \bar S+o\left({1\over\L_0^2}\right)\ .
\end{eqnarray}
Sending now $\L_0$ to infinity, we get a finite effective
superpotential of Veneziano-Yankielowicz type.\footnote{Of course,
to get the form of \cite{CIV01} the $-2\pi i\tau\bar S$-term can
be absorbed by redefining $\L=\left(\tilde\L
e^{-{2\pi i\tau}\over 3N}\right)^{3/2}$.}\\

\section{Conclusions}
\setcounter{equation}{0}
In this note we analysed the special
geometry relations on local Calabi-Yau manifolds of the form
\begin{equation}
W'(x)^2+f_0(x)+v^2+w^2+z^2=0\ .\label{locCYconc}
\end{equation}
The space of compact and non-compact three-cycles on this manifold
maps to the relative homology group $H_1(\S,\{Q,Q'\})$ on a
Riemann surface $\S$, given by $y^2=W'(x)^2+f_0(x)$, with two
marked points $Q,Q'$. We have shown that it is useful to split the
elements of this set into a set of compact cycles $\a^i$ and
$\b_i$ and a set containing the compact cycle $\hat\a$ and the
non-compact cycle $\hat\b$ which together form a symplectic basis.
The corresponding three-cycles on the Calabi-Yau manifold are
$\G_{\a^i},\G_{\b_j},\G_{\hat\a},\G_{\hat\b}$. This choice of
cycles is appropriate since the properties that arise from the
non-compactness of the manifold are then captured entirely by the
integral of the holomorphic three-form $\O$ over the non-compact
three-cycle $\G_{\hat\b}$ which corresponds to $\hat\b$. Indeed,
one finds the following relations
\begin{eqnarray}
-{1\over2\pi i}\int_{\G_{\a^i}}\O&=&2\pi i\tilde S_i\ ,\label{intalpha}\\
-{1\over2\pi i}\int_{\G_{\b_i}}\O&=&{\partial\mathcal{F}_0(t,\tilde S)\over\partial\tilde S_i}\ ,\label{intbeta}\\
-{1\over2\pi i}\int_{\G_{\hat\a}}\O&=&2\pi it\ ,\\
-{1\over2\pi i}\int_{\G_{\hat\b}}\O&=&
{\partial\mathcal{F}_0(t,\tilde S)\over\partial t}+W(\L_0)-t\log
\L_0^2+o\left({1\over\L_0}\right)\ .\label{intbetahat}
\end{eqnarray}
In the last relation the integral is understood to be over the
regulated cycle $\G_{\hat\b}$ which is an $S^2$-fibration over a
line segment running from the $n$-th cut to the cut-off $\L_0$.
Clearly, once the cut-off is removed, the last integral diverges.
To get rid of the polynomial divergence we introduced a pairing on
(\ref{locCYconc}) defined as
\begin{equation}\label{pairCY}
\left\langle\G_{\hat\b},\O\right\rangle:=\int_{\G_{\hat\b}}\left(\O-\d\Phi\right)=
(-i\pi)\int_{\hat\b}\left(\zeta-\d\varphi\right)\ ,
\end{equation}
where
\begin{equation}
\Phi:={W(x)W'(x)-{2t\over{n+1}}x^n\over W'(x)^2+f_0(x)}\cdot{\d
v\w\d w\over 2z}
\end{equation}
is such that
$\int_{\G_{\hat\b}}\d\Phi=-i\pi\int_{\hat\b}\d\varphi$. This
pairing is very similar in structure to the one appearing in the
context of relative (co-)homology and we proposed that one should
use this pairing so that Eq.(\ref{intbetahat}) is replaced by
\begin{equation}
-{1\over2\pi
i}\left\langle\G_{\hat\b},\O\right\rangle={\partial\mathcal{F}_0(t,\tilde
S)\over\partial t}-t\log \L_0^2+o\left({1\over\L_0}\right)\ .
\end{equation}
At any rate, whether one uses this pairing or not, the integral
over the non-compact cycle $\G_{\hat\b}$ is {\it not} just given
by the derivative of the prepotential with respect to $t$.\\
The set of cycles $\{\a^i,\hat\a,\b_i,\hat\b\}$ is particularly
convenient since we can perform arbitrary symplectic (duality)
transformations in $\{\a^i,\b_j\}$ without changing the structure
of the special geometry relations (\ref{intalpha}),
(\ref{intbeta}). However, once we mix $\b_i$- and $\hat\b$-cycles,
more special geometry relations are modified by cut-off dependent
terms.

Furthermore, we reconsidered the effective superpotential that
arises if we compactify IIB string theory on (\ref{locCYconc}) in
the presence of a background flux $H$. We emphasize that, although
the commonly used formula $W_{eff}=\int\O\w H$ is very elegant, it
should rather be considered as a mnemonic for
\begin{equation}\label{Wconc}
W_{eff}=\sum_{i=1}^{n-1}\left(\int_{\G_{\a^i}}H\int_{\G_{\b_i}}\O-\int_{\G_{\b_i}}
H\int_{\G_{\a^i}}\O\right)+\left(\int_{\G_{\hat\a}}H\left\langle\G_{\hat\b},
\O\right\rangle-\left\langle\G_{\hat\b},H\right\rangle\int_{\G_{\hat\a}}\O\right)
\end{equation}
because the Riemann bilinear relations do not necessarily hold on
non-compact Calabi-Yau manifolds. We noted that Eq.(\ref{Wconc})
is invariant under symplectic transformations of the basis of the
(relative) 3-cycles, provided one uses the pairing (\ref{pairCY})
whenever a relative cycle appears. Some of these transformations
act as electric-magnetic duality in the $U(1)^n$ gauge theory. By
manipulating (\ref{Wconc}) one obtains both the explicit results
of \cite{CIV01} and the more formal ones of \cite{DV02}. Although
the introduction of the pairing did not render the integrals of
$\O$ and $H$ over the $\G_{\hat\b}$-cycle finite since they are
still logarithmically divergent, these divergences cancel in
(\ref{Wconc}) and the effective superpotential is well-defined.

To derive these results we used the holomorphic matrix model as a
technical tool to find the explicit form of the prepotential. On
the way we have clarified several points related to the saddle
point expansion of the holomorphic matrix model. We showed that
although the partition function is independent of the choice of
the path $\g$ appearing in the matrix model, one has to choose a
specific path once one wants to evaluate the free energy from a
saddle point expansion. Since the spectral density $\r_0(s)$ of
the holomorphic matrix model is real by definition we found that
the cuts that form around the critical points of the
superpotential $W$ have specific orientations given by the second
derivatives $W''$ at the critical points. A path $\g$ that is
consistent with the saddle point expansion has then to be chosen
in such a way that all the cuts lie on $\g$. This guarantees that
one expands around a configuration for which the first derivatives
of the effective action indeed vanish. To ensure that saddle
points are really stable we were led to choose $\g$ to consist of
$n$ pieces where each piece contains one cut and runs from
infinity in one convergence domain to infinity of another domain.
Then the ``one-loop" term is a convergent, subleading Gaussian
integral. Using these results for the saddle point expansion of
the matrix model we then determined the free energy of the model
in the planar limit $\mathcal{F}_0(t,\tilde S_i)$. Here the
$\tilde S_i$ fix the fraction of eigenvalues that sit close to the
$i$-th critical point. The Riemann surfaces that appear in the
planar limit of the matrix model only are a subset of the more
general surfaces one obtains from the local Calabi-Yau manifolds,
since the $\tilde S_i$ are manifestly real. We proved the
(modified) special geometry relations in terms of
$\mathcal{F}_0(t,\tilde S_i)$ for these Riemann surfaces. These
relations can then be ``analytically continued" to complex values
of $t$ and $\tilde S_i$, and we used the same
$\mathcal{F}_0(t,\tilde S_i)$ to prove the modified special
geometry relations for the general hyperelliptic Riemann surface
(\ref{Riemannsurface}). One should note, however, that once $t$
and $\tilde S_i$ are taken to be complex, $\mathcal{F}_0(t,\tilde
S_i)$ still is the prepotential but it loses its interpretation as
the planar limit of the free energy of a a matrix model.

\vskip 17.mm \noindent {\bf\large Acknowledgements} \vskip 3.mm

\noindent Steffen Metzger gratefully acknowledges support by the
Gottlieb Daimler- und Karl Benz-Stiftung as well as by the
Studienstiftung des deutschen Volkes. We would like to thank Jan
Troost and Volodya Kazakov for helpful discussions.

\vskip 2.cm


\begin{thebibliography}{99}

\bibitem{CHSW85}
P. Candelas , G.T. Horowitz, A. Strominger and E. Witten, {\it
Vacuum configurations for superstrings}, Nucl. Phys. {\bf B258},
(1985) 46

\bibitem{dWvP}
B. de Wit and A. Van Proeyen, {\it  Potentials and symmetries of
general gauged $\mathcal{N}=2$ supergravity - Yang-Mills models},
Nucl. Phys. {\bf B245} (1984) 89; B. de Wit, P. Lauwers and A. Van
Proeyen, {\it Lagrangians of $\mathcal{N}=2$ supergravity - matter
systems}, Nucl. Phys. {\bf B255} (1985) 569; E. Cremmer, C.
Kounnas, A. Van Proeyen, J.P. Derendinger, S. Ferrara, B. de Wit
and L. Girardello, {\it Vector multiplets coupled to
$\mathcal{N}=2$ supergravity: superhiggs effect, flat potentials
and geometric structure}, Nucl. Phys. {\bf B250} (1985) 385

\bibitem{CO90}
P. Candelas and X. de la Ossa, {\it Moduli space of Calabi-Yau
manifolds}, Nucl. Phys. {\bf B355} (1991) 455

\bibitem{SW94}
N. Seiberg and E. Witten, {\it Electric-Magnetic Duality, Monopole
Condensation, and Confinement in $\mathcal{N}=2$ Supersymmetric
Yang-Mills Theory}, Nucl. Phys. {\bf B426} (1994) 19,
Erratum-ibid. {\bf B430} (1994) 485, {\tt hep-th/9407087}; {\it
Monopoles, Duality and Chiral Symmetry Breaking in $\mathcal{N}=2$
Supersymmetric QCD}, Nucl. Phys. {\bf B431} (1994) 484, {\tt
hep-th/9408099}

\bibitem{KLMVW96}
A. Klemm, W. Lerche, P. Mayr, C. Vafa and N. Warner, {\it
Self-Dual Strings and $\mathcal{N}=2$ Supersymmetric Field
Theory}, Nucl. Phys. {\bf B477} (1996) 746, {\tt hep-th/9604034}

\bibitem{KKV96}
S. Katz, A. Klemm and C. Vafa, {\it Geometric Engineering of Quantum Field Theories}, Nucl. Phys. {\bf B497} (1997)
173, {\tt hep-th/9609239}

\bibitem{AKMV03}
M. Aganagic, A. Klemm, M. Mari\~{n}o and C. Vafa, {\it The
Topological Vertex}, Commun. Math. Phys. {\bf 254} (2005) 425,
{\tt hep-th/0305132}

\bibitem{ADKMV03}
M. Aganagic, R. Dijkgraaf, A. Klemm, M. Mari\~{n}o and C. Vafa, {\it Topological Strings and Integrable
Hierarchies}, {\tt hep-th/0312085}

\bibitem{M04}
M. Mari\~no, {\it Chern-Simons Theory and Topological Strings}, {\tt hep-th/0406005}

\bibitem{KKLM99}
S. Kachru, S. Katz, A. Lawrence, J. McGreevy, {\it Open string
instantons and superpotentials}, Phys. Rev. {\bf D62} (2000)
026001, {\tt hep-th/9912151}

\bibitem{CIV01}
F. Cachazo, K.A. Intriligator and C. Vafa, {\it A Large $N$
Duality via a Geometric Transition}, Nucl. Phys. {\bf B603} (2001)
3, {\tt hep-th/0103067}

\bibitem{GV98}
R. Gopakumar and C. Vafa, {\it M-theory and topological
strings-I}, {\tt hep-th/9809187}; {\it M-theory and topological
strings-II}, {\tt hep-th/9812127}; {\it On the gauge
theory/geometry correspondence}, Adv. Theor. Math. Phys. {\bf 3}
(1999) 1415, {\tt hep-th/9811131}

\bibitem{DOV04}
U.H. Danielsson, M.E. Olsson and M. Vonk, {\it Matrix models, 4D
black holes and topological strings on non-compact Calabi-Yau
manifolds}, JHEP {\bf 0411} (2004) 007, {\tt hep-th/0410141}

\bibitem{V01}
C. Vafa, {\it Superstrings and topological strings at large $N$},
J. Math. Phys. {\bf 42}, (2001) 2798, {\tt hep-th/0008142}

\bibitem{DV02}
R. Dijkgraaf and C. Vafa, {\it Matrix Models, Topological Strings, and Supersymmetric Gauge Theories}, Nucl. Phys.
{\bf B644} (2002) 3, {\tt hep-th/0206255}; {\it On Geometry and Matrix Models}, Nucl. Phys. {\bf B644} (2002) 21,
{\tt hep-th/0207106}; {\it A Perturbative Window into Non-Perturbative Physics}, {\tt hep-th/0208048}

\bibitem{Wi95}
E. Witten, {\it Chern-Simons Gauge Theory as a String Theory},
Prog. Math. {\bf 133} (1995) 637, {\tt hep-th/9207094}

\bibitem{La03}
C.I. Lazaroiu, {\it Holomorphic matrix models}, JHEP {\bf 0305}
(2003) 044, {\tt hep-th/0303008}

\bibitem{AGLV}
V.I. Arnold et. al. {\it Singularity Theory}, Springer-Verlag,
Berlin Heidelberg 1998

\bibitem{L96}
W. Lerche, {\it Introduction to Seiberg-Witten Theory and its
Stringy Origin}, {\tt hep-th/9611190}

\bibitem{K99}
I.K. Kostov, {\it Conformal Field Theory Techniques in Random
Matrix Models}, {\tt hep-th/9907060}

\bibitem{K91}
I.R. Klebanov, {\it String theory in two dimensions}, Trieste
Spring School (1991) 30, {\tt hep-th/9108019}

\bibitem{CRTP97}
B. Craps, F. Roose, W. Troost and A. Van Proyen, {\it What is
Special K\"ahler Geometry?}, Nucl. Phys. {\bf B503} (1997) 565,
{\tt hep-th/9703082}

\bibitem{KL87}
M. Karoubi and C. Leruste, {\it Algebraic Topology via
Differential Geometry}, Cambridge University Press, Cambridge 1987

\bibitem{FK}
H.M. Farkas and I. Kra, {\it Riemann Surfaces}, Springer Verlag,
New York 1992

\bibitem{LMW02}
W. Lerche, P. Mayr and N. Warner,  {\it Holomorphic N=1 Special
Geometry of Open-Closed Type II Strings}, {\tt hep-th/0207259};
{\it N=1 Special Geometry, Mixed Hodge Variations and Toric
Geometry}, {\tt hep-th/0208039}; W. Lerche, {\it Special Geometry
and Mirror Symmetry for Open String Backgrounds with N=1
Supersymmetry}, {\tt hep-th/0312326}



\end{thebibliography}
\end{document}